\documentclass[preprint]{aastex}

\slugcomment{Submitted to AJ}
\shorttitle{Formation of Uranus and Neptune}
\shortauthors{Thommes et al.}
\begin{document}
\title{The formation of Uranus and Neptune among Jupiter and Saturn}
\author{E. W. Thommes}
\affil{Astronomy Department, University of California,
    Berkeley, CA 94720} 
\email{ethommes@astro.berkeley.edu}

\author{M. J. Duncan}
\affil{Physics Department, Queen's University, Kingston, Ontario K7L 3N6}
\email{duncan@astro.queensu.ca}

\author{H. F. Levison}
\affil{Southwest Research Institute, Boulder, CO 80302}
\email{hal@gort.boulder.swri.edu}

\begin{abstract}
The outer giant planets, Uranus and Neptune, pose a challenge to
theories of planet formation.  They exist in a region of the Solar
System where long dynamical timescales and a low primordial density of
material would have conspired to make the formation of such large
bodies ($\sim 15$ and 17 times as massive as the Earth, respectively)
very difficult.  Previously, we proposed a model which addresses this
problem: Instead of forming in the trans-Saturnian region, Uranus and
Neptune underwent most of their growth among proto-Jupiter and
-Saturn, were scattered outward when Jupiter acquired its massive gas
envelope, and subsequently evolved toward their present orbits.  We
present the results of additional numerical simulations, which further
demonstrate that the model readily produces analogues to our Solar
System for a wide range of initial conditions.  We also find that this
mechanism may partly account for the high orbital inclinations
observed in the Kuiper belt.

\end{abstract}


\section{Introduction}
\label{introduction}

The growth of Uranus and Neptune in the outer Solar System is not
readily accounted for by conventional models of planet formation.  A
low primordial density of planetesimals and weak solar gravity would
have made the process of accretion slow and inefficient.  In direct
N-body simulations of accretion among (approximately) Earth-mass bodies beyond
10 AU, performed with three different computer codes, little accretion
is found to take place over timescales of $10^8$ years, and by
extrapolation, over the age of the Solar System (Levison \& Stewart
2001).  Earlier simulations by Brunini \& Fernandez (1999) showed
accretion of the ice giants in several $\times 10^7$ years with the
same initial conditions, but later simulations, performed with an
improved integrator, require that bodies be enhanced in radius by at
least a factor of ten, relative to bodies having the density of Uranus
and Neptune, in order to recover the previous result (Brunini 2000).
Therefore, Uranus and Neptune are unlikely to have formed from a late
stage of mergers among large protoplanets, analogous to the putative
final phase of planet formation in the terrestrial zone (eg. Wetherill
1996, Chambers \& Wetherill 1998).  The oligarchic growth model, in
which the principal growth mode is accretion of small planetesimals by
a protoplanet, also produces timescales which are too long (Kokubo and
Ida 2000, Thommes 2000), unless the feeding zones of Uranus and
Neptune can be replenished quickly enough with low random velocity
planetesimals from elsewhere in the nebula (Bryden, Lin and Ida 2000).

Thommes, Duncan \& Levison (1999), hereafter TDL99, develop an
alternative model to in-situ formation for the origin of Uranus and
Neptune.  Beginning with four or more planetary embryos of 10-15
M$_{\oplus}$ in the Jupiter-Saturn region, they explore through N-body
simulation the evolution of the system after one of these bodies (and
in one case, two at the same time) accretes a massive gas envelope to
become a gas giant.  They find that the remaining giant protoplanets
are predominantly scattered outward.  Dynamical friction with the
planetesimal disk subsequently recircularizes their orbits, which
leads, in about half the simulations performed, to a configuration
quite similar to the present outer Solar System, with the scattered
giant protoplanets taking the roles of Uranus, Neptune and Saturn.
These results suggest that Uranus and Neptune are actually potential
gas giant cores which formed in the same region as Jupiter and Saturn,
but lost the race to reach runaway gas accretion.

Here, we explore this model in more detail, and perform further
simulations.  Sections \ref{available_material} and \ref{oligarchic
implications} motivate our choice of initial conditions for the
simulations.  Section \ref{scatter+circ} discusses the mechanism for
transporting a proto-Uranus and -Neptune to the outer Solar System.
N-body simulation results are presented in Section \ref{n-body
simulations}.  The effect on the asteroid and Kuiper belts is
discussed in Section \ref{small body belts}.  We summarize and discuss
our findings in Section \ref{discussion}.

\section{Available material in the Jupiter-Saturn region}
\label{available_material}

Hayashi (1981) estimated the minimum primordial surface density of
solids in the outer Solar System to be

\begin{equation}
\sigma_{min}(a) = 2.7(a/5\,AU)^{-3/2}\,g/cm^2.
\label{minimum_mass_solids}
\end{equation}
  
The requirement that Jupiter and Saturn's cores be massive enough to
have initiated runaway gas accretion suggests that they are $\sim$ 10
M$_{\oplus}$ in mass (Mizuno et al 1978, Pollack et al 1996).
Interior models are consistent with such a core mass, but also allow a
coreless Jupiter (Guillot 1999).  We assume in this work that both
Jupiter and Saturn began as $\sim$ 10 M$_{\oplus}$ bodies; putting gas
giant cores and ice giant cores on the same footing is necessary in
the ``strong'' version of our model, though it is not essential to the
basic mechanism; we discuss variations on the model in Section
\ref{discussion}.

A surface density of 2.7 g/cm$^2$ was likely too low to form a $\sim$
10 M$_{\oplus}$ body at 5 AU before the gas was removed from the
protoplanetary disk.  Lissauer (1987) finds that a surface density of
15-30 g/cm$^2$ is needed to allow formation of Jupiter's core on a
timescale of $5\times 10^5 - 10^6$ years, while the model of Pollack
et al (1996), which includes concurrent accretion of solids and gas,
produces Jupiter in less than $10^7$ years with 10 g/cm$^2$.  The
formation of giant planet cores may have been triggered at least in
part by the enhancement in the solids surface density beyond the
``snow line'', where water goes from being a gaseous to a solid
constituent of the protoplanetary disk.  In fact, outward diffusion
and subsequent freezing of water vapor from the inner Solar System may
have resulted in a large local density enhancement around 5 AU,
perhaps yielding a surface density even higher than 30 g/cm$^2$
(Stevenson \& Lunine 1988).  Here we assume a power law surface
density,
\begin{equation}
\sigma(a) = \sigma_0 (a/5\,AU)^{-\alpha}.
\end{equation}
The above discussion suggests $\sigma \sim$ 10 to 30 g/cm$^2$ as a
plausible surface density at 5 AU.  Allowing, also, the exponent
$\alpha$ to vary between 1 and 2, one obtains a total mass in the
Jupiter-Saturn (J-S) region in excess of 40 M$_{\oplus}$, and as high
as 180 M$_{\oplus}$.  Therefore, it is likely that the region
originally contained significantly more solids than ended up in the
cores of Jupiter and Saturn.

\section{Implications of oligarchic growth}
\label{oligarchic implications}

Runaway growth (eg. Wetherill \& Stewart 1989, Kokubo \& Ida
1996) transitions to a slower, self-limiting mode called ``oligarchic
growth'' (Ida \& Makino 1993, Kokubo \& Ida 1998, 2000) when the
largest protoplanets are still orders of magnitude less than an Earth
mass everywhere in the nebula.
Oligarchic growth has previously only been demonstrated to take place
interior to about 3 AU (Weidenschilling \& Davis 2000).  Though
Kokubo and Ida make estimates of protoplanet mass and growth
timescales in the giant planet region, they point out that their
simulations are restricted to annuli which are narrow compared to
their radii, and thus cannot make strong predictions about how
oligarchic growth works over a wide range in semimajor axis (Kokubo
and Ida 2000).  However, their most recent simulations (Kokubo \& Ida
2000b) span a range of 0.5 to 1.5 AU, and show oligarchic growth
proceeding in an outward-expanding ``wave'' over time.  Also, Thommes
(2000) and Thommes, Duncan \& Levison (2001) perform numerical
simulations which suggest that oligarchic growth does take place in
the outer Solar System, and proceeds on approximately the timescales
predicted using the approach of Kokubo \& Ida (2000).

Assuming that protoplanets remain approximately evenly spaced in Hill
radii while growing---as is characteristic of oligarchic
growth---their final mass is given by
\begin{equation}
M = (2 \pi)^{3/2} \left ( \frac{2}{3 M_{\odot}} \right )^{1/2} p^{3/2}
n^{3/2} \sigma^{3/2} a^3
\label{oligarchic_mass}
\end{equation}
where $a$ is the protoplanet semimajor axis, $n$ is the spacing
between adjacent protoplanets in Hill radii $r_H$, and $p$ is the
fraction of the total mass in the zone $[a-n r_H/2, a+n r_H/2]$
incorporated into the protoplanet (Kokubo and Ida 2000).  Using
numerical simulations, Kokubo \& Ida (1998) show that $n \sim 5 - 10$.
Models of giant planet formation by concurrent planetesimal and gas
accretion suggest a surface density profile $\sigma \propto a^{-2}$
(Pollack et et 1996).  Using this in Eq. \ref{oligarchic_mass}, one
obtains a protoplanet mass independent of semimajor axis.  Adopting
$\sigma_0 = 10\,g/cm^2$ and a spacing of n=7.5 Hill radii, one must
then set the accreted mass fraction to about 0.8 in order to obtain a
protoplanet mass of 10 M$_{\oplus}$ .  With this spacing, between
three and five such bodies fit between 5 and 10 AU.  It is likely,
therefore, that this region originally contained more than just the
future solid cores of Jupiter and Saturn.  At the same time, the more
recent simulations of Thommes, Duncan and Levison (2001, preprint)
indicate that accretion is quite inefficient in the outer Solar
System; even in the Jupiter-Saturn region, the above value of $p$ is
probably overly optimistic, and a correspondingly higher density of
solids in this region may have been necessary to form gas giant cores.
But as we find (Section \ref{set_2}), a higher-density disk actually
tends to increase the ``success rate'' of the model.

\section{Scattering and circularization of the protoplanets}
\label{scatter+circ}
In the nucleated instability picture of gas giant formation, there are
three distinct phases of growth (Pollack et al 1996): In Phase 1
(which itself encompasses the sub-phases of runaway and oligarchic
growth; see Section \ref{oligarchic implications} above), a solid core
grows until it depletes most of the material in its feeding zone.
Phase 2 is characterized by much slower growth, with gas accretion
gradually coming to dominate over solids accretion.  After a time of
order several Myrs, the protoplanet contains comparable masses of gas
and solids; around this time, the third phase of runaway gas accretion
sets in and proceeds on a timescale of $\sim 10^5$ years.  One of the
protoplanets in the Jupiter-Saturn region must have been the first to
reach this point.  Shorter formation timescales at smaller
heliocentric radii argue for Jupiter having formed before Saturn.  The
relatively long time spent by a protoplanet on the ``plateau'' of
Phase 2 means that several protoplanets could plausibly find
themselves there by the time the first of them makes it to Phase 3.
Furthermore, if the winner has a broad enough margin of victory, its
rivals will have only had time to accumulate a few M$_{\oplus}$ of
gas, thus resembling the present-day ice giants in both solids and gas
mass.
 
A body increasing its mass from 10 M$_{\oplus}$ to Jupiter's mass,
M$_J$ = 314 M$_{\oplus}$, expands its Hill radius by a factor of about
three.  The adjacent protoplanets will therefore have a high
likelihood of being gravitationally scattered.  In a three-body
system---Sun, Jupiter and a single protoplanet---a Jupiter-crossing
protoplanet would continue to have close encounters with its
much more massive scatterer, and so would remain coupled to it, unless
it were scattered onto an unbound orbit and left the system
altogether.
In reality, however, one expects that as it crosses beyond the
Jupiter-Saturn region, a body will encounter a less accretionally
evolved part of the protoplanetary disk, consisting predominantly of
much smaller bodies.  As a result, the protoplanet will experience
dynamical friction.  This will tend to reduce the eccentricity of its
orbit.  If the eccentricity decays enough, the protoplanet's
perihelion will be lifted away from Jupiter, thus decoupling it from
its scatterer.  Insofar as one can neglect interactions with other
scattered protoplanets, its eccentricity will then monotonically
decrease until it reaches equilibrium with the planetesimals.  The
eccentricity rate of change due to dynamical friction on a body of
mass $M$ in a swarm of mass $m$ bodies can be expressed as
(Weidenschilling et al 1997)
\begin{equation}
\frac{d e^2_M}{dt}=C(m \langle e_m^2 \rangle - M
e_M^2)K_e
\label{dynamical_friction_ecc}
\end{equation}
and a similar expression gives the time evolution of the inclination.
$\langle e_m^2 \rangle^{1/2}$ is the RMS eccentricity of the mass $m$
bodies, and $K_e$ is a definite integral depending on the inclinations
and eccentricities, defined in Stewart \& Wetherill (1988).
The coefficient C is given by
\begin{equation}
C=\frac{16 G^2 {\rho_{sw}}_1 ln \Lambda}{v^3_K (\langle e_1^2 \rangle +
\langle e_2^2 \rangle)^{3/2}},
\label{dynamical_friction_coeff}
\end{equation}
where $G$ is the gravitational constant, $v_K$ the local Keplerian
velocity, ${\rho_{sw}}_1$ the spatial mass density of the swarm, and
$\Lambda$ is approximately the ratio of the maximum encounter distance
between the body $M$ and a member of the population of the swarm, to
the maximum separation that results in a physical encounter (Stewart
\& Wetherill 1988).  

An equilibrium is therefore reached when
\begin{equation}
e_M^2 M \simeq \langle e_m^2 \rangle m
\label{dynamical_friction_equilibrium}
\end{equation}
This also means that as long as 
\begin{equation}
e_M^2 M \gg \langle e_m^2 \rangle m,
\label{mass_indep_condition}
\end{equation} 
the eccentricity decay rate of the body $M$ will only depend on the
spatial mass density of the planetesimal disk, and will be essentially
independent of the individual masses of the planetesimals.  This
feature will be exploited in the numerical simulations below.  

\section{N-body simulations} 
\label{n-body simulations}

Putting the pieces together, one can now envision a scenario in which
a) a rapidly growing Jupiter scatters its smaller neigbours outward,
and b) these ``failed cores'' are decoupled from Jupiter and
ultimately evolve onto circular, low-inclination orbits in the outer
Solar System.  TDL99 performed three series of numerical simulations
to test this model; about half of the simulations produced a stable
final configuration qualitatively similar to the present-day Solar
System.  Here, we expand on the earlier work, performing simulations
to assess the effect of other planetesimal disk density profiles and
of proto-Jupiter being initially not the innermost giant protoplanet.
We also investigate how sensitive this scenario is to the timing of
Saturn's final stage of growth, relative to that of Jupiter.
Simulations are once again performed using the proven SyMBA symplectic
integrator (Duncan, Levison \& Lee 1998).  This integrator is able to
handle close encounters among massive bodies while preserving the
symplectic properties of the method of Wisdom \& Holman (1991).

\begin{figure}
\begin{center}
\includegraphics[width=5.0in]{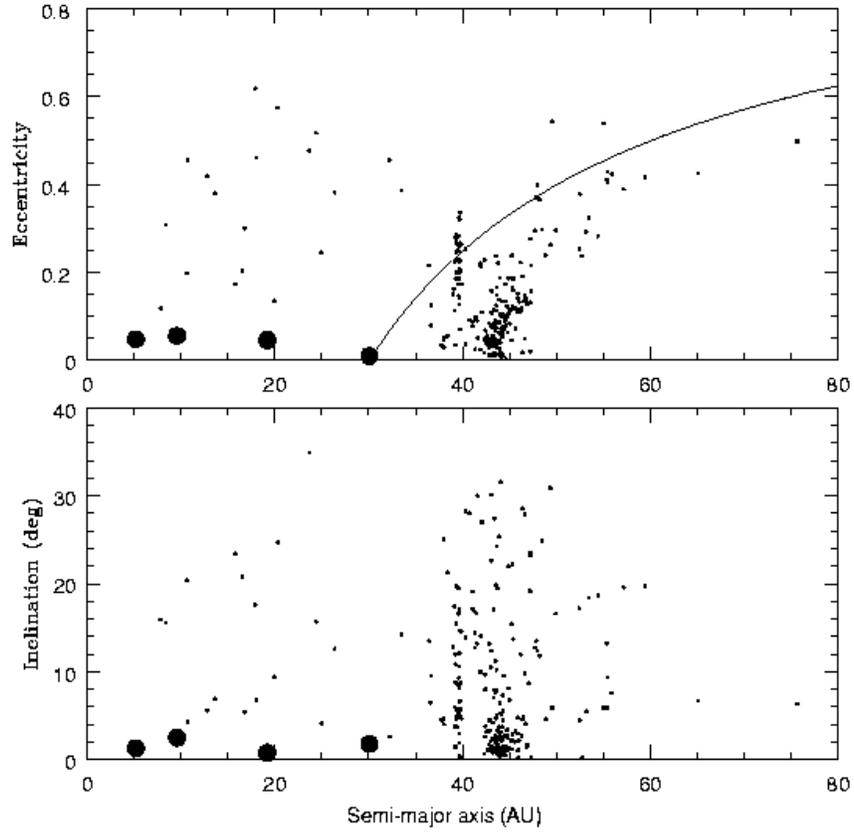}
\caption{Eccentricities (top) and inclinations (bottom) in the outer
Solar System at the present epoch, showing the giant planets as well
as all Kuiper belt objects and Centaurs (objects with a $\la$
30 AU) which have been observed at multiple oppositions, as of
September 2001.  The curve in the top panel shows the
locus of orbits with perihelia at the semimajor axis of Neptune.
KBO and Centaur data is taken from the Minor Planet Center site,
cfa-www.harvard.edu/iau/mpc.html }
\label{real_solar_system}
\end{center}
\end{figure}

\subsection{Initial conditions: Set 1}
\label{set 1}
For the baseline set of simulations, a planetesimal disk of surface
density 
\begin{equation}
\sigma_1 = 10(a/5\,AU)^{-2}\,g/cm^2
\label{density_set1}
\end{equation}
is used, with the disk extending from 5 to 60 AU.  The above density
profile is steeper than those of TDL99 ($\sigma \propto a^{-1}$ and
$a^{-3/2}$).  The total disk mass between 10 and 60 AU (106
M$_{\oplus}$) is about half that of the first two sets of runs in our
earlier work (216 M$_{\oplus}$), and similar to that in the third run
(119 M$_{\oplus}$).  The truncation at 60 AU is to keep the number of
bodies in the simulations tractably small; in reality the planetesimal
disk may have extended for hundreds of AU, far beyond the presently
visible Kuiper belt.  Observational evidence does point to the
possibility of a truncation of the belt at $\sim$ 50 AU (eg. Chiang \&
Brown 1999).  But if a ``Kuiper cliff'' exists, it is not necessarily
primordial.

Since we truncate the disk at a location where the planetesimal
surface density is already very low (0.07 g/cm$2$ for the density
profile of Eq. \ref{density_set1}), one expects this not to have a
strong effect on the evolution of any scattered protoplanets which
cross the region $r > 60$ AU.  As a test, we perform four pairs of
runs comparing the evolution of a 10 M$_{\oplus}$ body with initial
$a=100$ AU, $e=0.9-0.95$ in planetesimal disks with a surface density
as prescribed by Eq. \ref{density_set1}, in one case extending to 60
AU, in the other to 200 AU.  No systematic difference in the 10
M$_{\oplus}$ body's semimajor axis evolution over the first few Myrs
is apparent between the two different disk sizes.  The eccentricity,
however, tends to be damped more rapidly in a 200 AU disk.  We expect,
therefore, that the subsequent runs somewhat underestimate the
effectiveness with which those scattered protoplanets which cross
beyond 60 AU are circularized, if the disk was in fact larger.  In
particular, a planetesimal disk with a radius of hundreds of AU may
allow the retention of protoplanets which in our runs are scattered
strongly enough to become unbound; we will explore this possibility in
future work.
 
The simulated disk is made up of equal-mass ``planetesimals'', each
having a mass of 0.2 M$_{\oplus}$.  At twice the mass of Mars, these
bodies far exceed the actual characteristic mass of planetesimals in
the early outer Solar System.  In reality planetesimals are thought to
have been on the order of 1 to 100 km in size, thus with a mass of
$\sim 10^{-12} - 10^{-6}$ M$_{\oplus}$ (eg. Lissauer 1987).  The
unrealistically large masses are chosen to keep the number of bodies
manageably low, at slightly over 500.  As mentioned in Section
\ref{scatter+circ}, when the large bodies have eccentricities high
enough that Eq. \ref{mass_indep_condition} is satisfied, the
eccentricity decay timescale will be effectively independent of the
planetesimal masses.  Thus despite the large planetesimal masses,
the initial strength of eccentricity damping will be realistic.  The
equilibrium eccentricity condition
Eq. \ref{dynamical_friction_equilibrium}, however, does depend on the
planetesimal mass, so the equilibrium eccentricity reached by a large
body among the planetesimals in the simulation will tend to be
unrealistically high.  Of course, a true equilibrium between
protoplanets and planetesimals will not exist anyway, since mutual
perturbations among the protoplanets will also have an effect.

The initial planetesimal eccentricities and inclinations are given a
Rayleigh distribution in $e$ and $i$, (Kokubo and Ida 1992) with
$\langle e^2 \rangle^{1/2} = 0.05$, $\langle i^2 \rangle^{1/2} = 0.025
= 1.4^o$.  In the numerical integrations, the planetesimals are
treated as a ``second-class'', non-self-interacting population.  Thus
they are perturbed in their Keplerian orbits only by forces from the
protoplanets, not each other.  The protoplanets, on the other hand,
are subject to forces from each other as well as from the
planetesimals.  This serves two purposes: It makes the simulations run
much faster, since for $N$ second-class bodies, the computation time
scales as $N$ instead of $N^2$.  Also, it prevents unrealistically
strong self-stirring of the disk.  Of course, not modeling
planetesimal interactions means that collective planetesimal effects
are not accounted for.  Wave phenomena could have had an important
effect on the evolution of the planetesimal disk velocity distribution
(eg. Ward and Hahn 1998), provided the disk was sufficiently massive
and dynamically cold.  However, it is unlikely that significant wave
phenomena could persist once the initial scattering has taken place
and the planetesimal disk has been stirred by eccentric 10
M$_{\oplus}$ bodies.

Also unmodeled is nebular gas, either as a source of aerodynamic drag
(relevant for small planetesimals; eg. Adachi, Hayashi and Nakazawa
1976), or as a source of tidal forces (relevant for bodies
$\ga$ 0.1 M$_{\oplus}$; eg. Ward 1997).  The former mechanism
will keep the planetesimal disk more dynamically cold, but this does
not affect the strength of dynamical friction on larger bodies until
$e^2_M M \sim e^2_m m$.  The latter effect is thought to cause the
inward migration of $\sim$ 1 - 10 M$_{\oplus}$ objects on timescales
short compared to their formation times.  This of course constitutes a
potentially severe problem not only for our model, but for any model
of giant planet formation which requires the accumulation of large
solid cores.  Addressing this problem is beyond the scope of our
present work, but we summarize some reasons why it may in reality have
been less severe, in Section \ref{discussion}.

The 10 M$_{\oplus}$ bodies are given a density $\rho = 0.25
\rho_{\oplus} = 1.5$ g/cm$^3$, roughly equal to that of Uranus and
Neptune.  This gives them radii of $2.18 \times 10^4$ km.  Four such
bodies are put in the simulation, initially on nearly circular and
uninclined orbits.  The orbits are spaced by 7.5 mutual Hill radii,
starting from an innermost distance of 6 AU to allow for later
inward migration by Jupiter.  Thus the bodies' initial semimajor axes
are 6.0 AU, 7.4 AU, 9.0 AU and 11.1 AU.  Between $5$ and
12 AU, the disk is depleted in planetesimals so that the surface
density is still given by Eq. \ref{density_set1}.  Since the large
bodies are spaced proportionally to their semimajor axes, their
distribution is consistent with a surface density $\propto a^{-2}$.

\begin{figure}
\begin{center}
\includegraphics[width=5.0in]{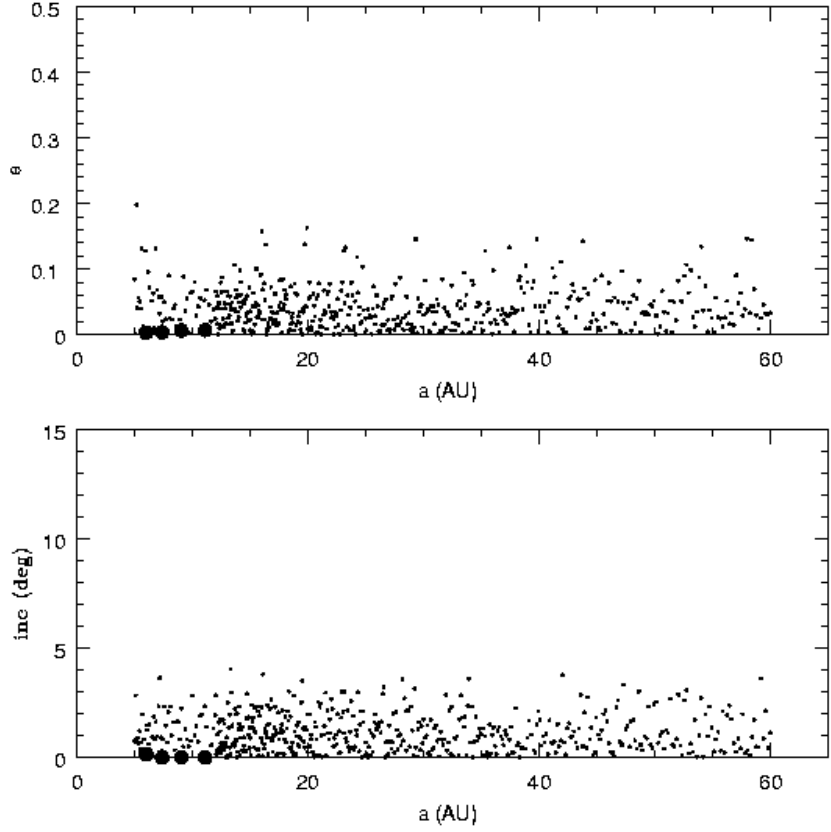}
\caption{Initial state for runs in Set 1, showing eccentricity
(top) and inclination (bottom) versus semimajor axis.  The larger
circles denote the four 10 M$_{\oplus}$ protoplanets, and each of the small
dots represents a 0.2 M$_{\oplus}$ planetesimal.  The
planetesimal density in the vicinity of the protoplanets is decreased to
keep the density of protoplanets plus planetesimals consistent with the
surface density given by Eq. \ref{density_set1}.}
\label{run21_initial_state}
\end{center}
\end{figure}

Integrating the surface density from $5$ to 12 AU, the total mass
is 51.5 M$_{\oplus}$.  With 40 M$_{\oplus}$ of this in the large
bodies, this leaves 10.5 M$_{\oplus}$ in planetesimals in the
Jupiter-Saturn region, in addition to 94.7 M$_{\oplus}$ between 12 AU,
and the outer disk edge at 60 AU.  In summary, then, the initial
conditions amount to a state where Phase 2 of giant planet formation
has been reached between 5 and 12 AU, with $\sim 80 \%$ of the
planetesimals having accreted through oligarchic growth into four
bodies of 10 M$_{\oplus}$ each, while no large bodies have yet formed
beyond this region (Fig. \ref{run21_initial_state}).  The adoption of
equal-mass protoplanets is a simplification; apart from stochastic
variations among the oligarchic growth endproducts, one can expect
some intermediate bodies, with masses perhaps up to a few
M$_{\oplus}$.  Such bodies are likely to ultimately be cleared, along
with the planetesimals, from the giant planet region.  However, they
may end up playing a role in the dynamics of the trans-Neptunian
region; see Petit, Morbidelli and Valsecchi (1999).

For computational reasons, the inner simulation radius is chosen as 1
AU; any body whose orbit penetrates this boundary is eliminated from
the system.  This is done because a limitation of the SyMBA integrator
used here is its inability to handle close perihelion passages.
Although a new version of SyMBA has since been developed which removes
this restriction (Levison and Duncan 2000), the runs presented here
predate this development.  In any case, this limitation has
little relevance for the runs presented here; typically, less than ten
planetesimals are lost at the inner boundary over the course of a 5
Myr simulation.  The base timestep is chosen as 0.05 years, giving 20
steps per orbital period for an orbit with its semimajor axis at the
inner radius, and over 200 steps per orbit for Jupiter.
Experimentation shows that this timestep is small enough that the
energy of the system is well-conserved.  Runs initially go to 5 Myrs;
in cases where the system still appears to be undergoing rapid
evolution at this point, the runs are extended by another 5 Myrs.

\subsection{Set 1 results}
\label{set_1}

To model gas accretion, SyMBA was modified to allow a subset of bodies
to have artificially time-varying masses (in addition to any changes
in mass resulting from the accretion of other bodies).  For the runs
in Subset 1, it is assumed that the innermost protoplanet undergoes
runaway gas accretion first, and grows into Jupiter.  The simulations
start at the time when this happens, which should be a few million
years into the life of the solar system (see above).  Runaway gas
accretion is simulated by increasing the body's mass over the first
$10^5$ years of simulation time, from its original mass of
10 M$_{\oplus}$ to 314 M$_{\oplus}$, approximately the present
Jupiter mass.  A linear growth in mass is used; this is deemed
appropriate since the actual time evolution of mass during runaway
growth is highly uncertain.  Also, as the simulations will show,
$10^5$ years is roughly the response time of the system, and the
system's subsequent evolution is therefore unlikely to be affected in
a systematic way by the exact form of the time evolution of the
runaway-phase mass growth.

Set 1 consists of eight alternate realizations of a run, differing
only in the initial phases of the four protoplanets; for each, the
angles $\Omega$, $\omega$ and $M$ \label{mean anomaly}are randomly
generated.  As will be seen, the stochasticity of the system ensures
that this difference in phases is sufficient to bring about a very
different evolution in each of the versions of the run.

\begin{figure}
\begin{center}
\includegraphics[width=5.0in]{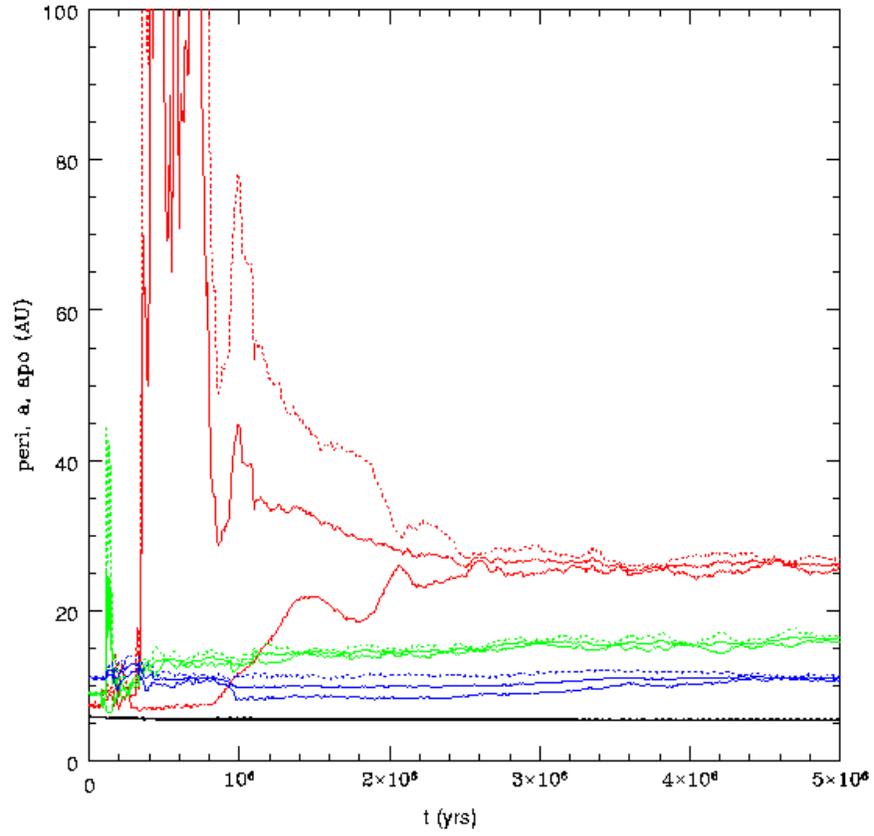}
\caption{Run 1F: Evolution of semimajor axis (bold lines), perihelion distance
$q$ (thin lines) and aphelion distance $Q$ (dotted lines) of the four
 10 M$_{\oplus}$ protoplanets.  The protoplanet which grows to Jupiter
mass (314 M$_{\oplus}$) over the first $10^5$ years of simulation
time is shown in black.}
\label{run21_a6_a_peri_apo}
\end{center}
\end{figure}

Fig. \ref{run21_a6_a_peri_apo} shows the evolution of one of the eight
runs, denoted as Run 1F, which after 5 Myrs produces final protoplanet
orbits that bear a particularly close resemblance to those of the
giant planets in the present-day Solar System.  Semimajor axis versus
time is plotted for each of the four protoplanets; the innermost one
has grown into Jupiter after the first $10^5$ years.  By this time,
the protoplanet orbits begin to mutually cross, and strong scattering
occurs.  The protoplanet plotted in red briefly has its semimajor axis
increased to greater than 100 AU.  However, dynamical friction acts to
reduce eccentricities universally, decoupling the protoplanets from
Jupiter and from each other.  By about 1.2 Myrs, none of the
protoplanets are on crossing orbits anymore.  After about 3 Myrs, the
bodies no longer undergo any changes in semimajor axis greater than a
few AU on a million-year timescale.  At this point the orbits are well
spaced and all eccentricities are $\la 0.05$, with no large
fluctuations.  Due to the large stochastic variations among runs, the
exact timescales differ, but we find that orbits typically become
noncrossing after less than 5 Myrs in these and subsequent runs.

\begin{figure}
\begin{center}
\includegraphics[width=5.0in]{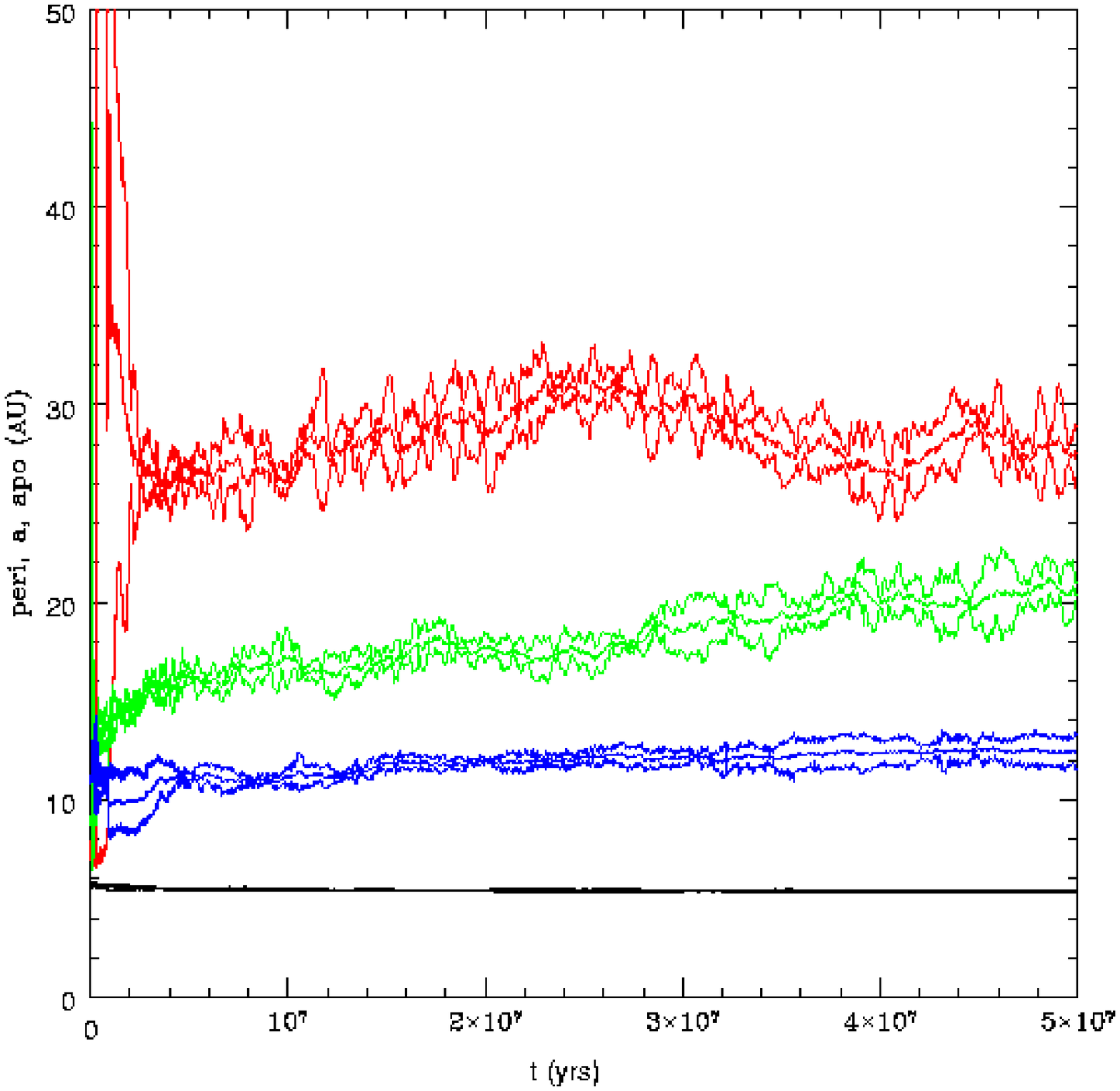}
\caption{Run 1F continued to 50 Myrs.  Between 5 Myrs and 50 Myrs, the
net migration for Jupiter and the protoplanets, going from inside to
outside in semimajor axis, is -0.2 AU, 1.5 AU, 4 AU, and
1.3 AU.}
\label{project11_run21_a6_resume_a_peri_apo}
\end{center}
\end{figure}

Subsequent semimajor axis evolution proceeds by scattering of
planetesimals by protoplanets, rather than scattering of the
protoplanets off each other and Jupiter.  As planetesimals are
scattered among Jupiter and the protoplanets, the former experiences a
net loss of angular momentum while the latter experience a gain.  Thus
Jupiter's orbit shrinks, while those of the protoplanets expand
(Fernandez and Ip 1996, Hahn and Malhotra 1999).  This phase takes
place over a timescale of several tens of Myrs and ends, at the very
latest, when the planetesimals have been cleared from among the
planets.  Because the length of migration during this phase is only a
few AU, and to save time, most of the runs are stopped after 5 Myrs.
As an example, Fig. \ref{project11_run21_a6_resume_a_peri_apo} shows
Run 1F continued to 50 Myrs; the net migration of the outer two
protoplanets subsequent to 5 Myrs is only $\sim$ 1 - 4 AU outward.

The semimajor axes at which the scattered bodies end up are very
noteworthy, if one compares them to the present orbits of the giant
planets (Fig. \ref{real_solar_system}).  At 5 Myrs, the outer two
protoplanets are at 16 and 26 AU.  The innermost one is at
11 AU, while Jupiter is at 5.5 AU.  This configuration of orbits
is very similar to that of the present Solar System, where Uranus and
Neptune are at 19 and 30 AU respectively, Saturn is at 9.6 AU,
and Jupiter is at 5.2 AU.  And at $5 \times 10^7$ years, after some
more net outward migration, the outer two protoplanets' semimajor axes
are even closer to those of Uranus and Neptune (though
``proto-Saturn'', having also moved outward, is further away from its
present orbit).  Eccentricities and inclinations are likewise very
close to their present values.

\begin{figure}
\begin{center}
\includegraphics[width=5.0in]{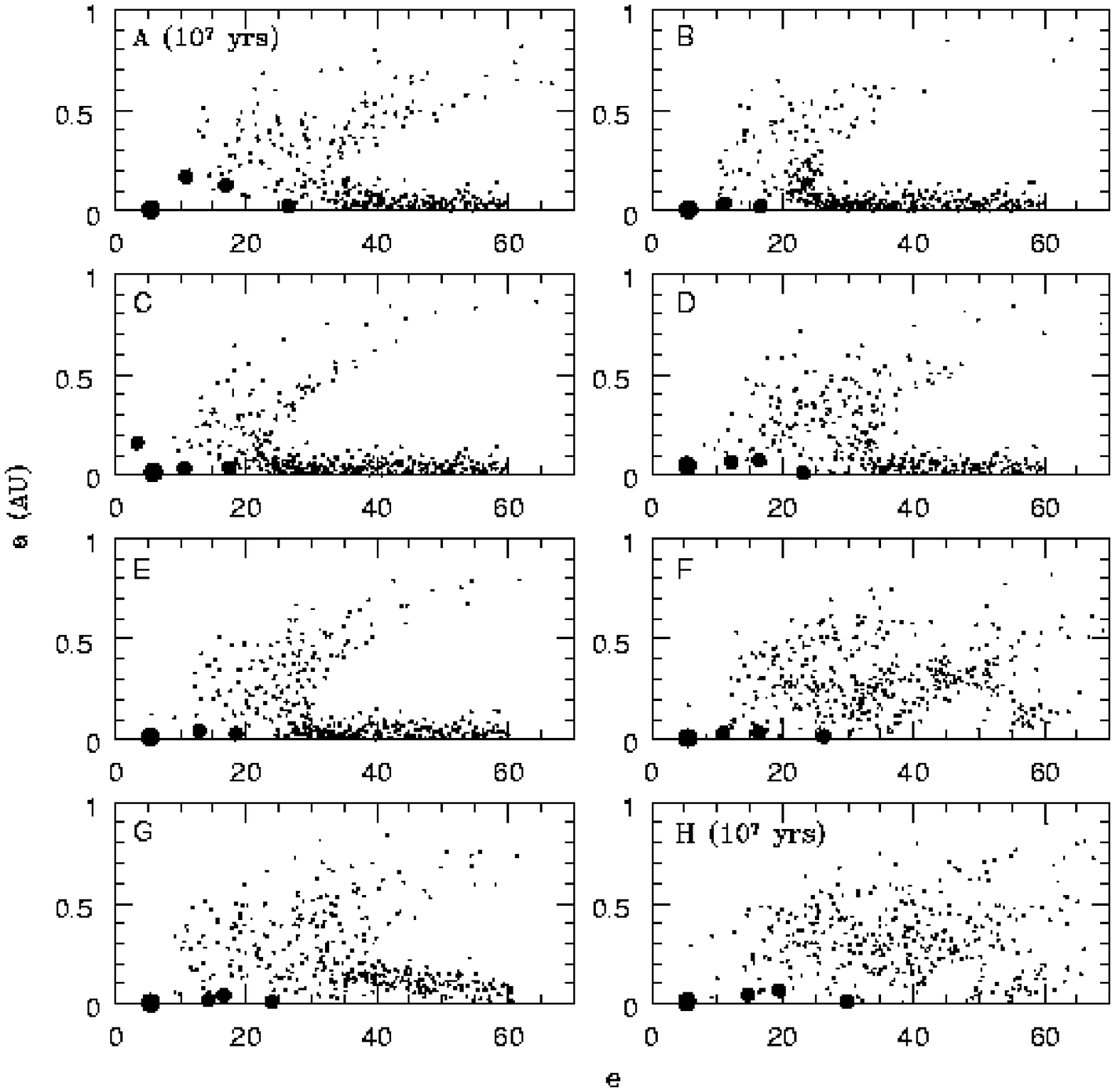}
\caption{Endstates of the eight Set 1 runs, after 5 Myrs of simulation
time, except for 1A and 1H, which were continued on to 10 Myrs.
Eccentricity is plotted versus semimajor axis.  The three different
sizes of points denote planetesimals (smallest), 10 M$_{\oplus}$ protoplanets (medium),
and Jupiter (largest).  Planetesimal orbits crossing Jupiter or any of the
protoplanets are generally unstable on timescales short compared to
the age of the Solar System, thus the region among the protoplanets
would be essentially cleared of planetesimals long before the present
epoch.}
\label{run21_a_ea_thesis}
\end{center}
\end{figure}
The end states of all the runs in Set 1 are summarized in
Fig. \ref{run21_a_ea_thesis}.  Depicted are snapshots of
eccentricity versus semimajor axis at 5 Myrs, except for Runs 1A and
1H, which were continued to 10 Myrs.  These were extended because one
or more of the protoplanets still had a high but decreasing
eccentricity at 5 Myrs.

Runs 1A, 1D, 1F, 1G and 1H result in a final ordering of orbits that
is at the very least broadly consistent with the present Solar System:
Jupiter is the innermost body, with the other three bodies interior to
the region of the Kuiper belt, and eccentricities low enough that no
protoplanets cross each other or Jupiter.  Out of these five runs, 1D
and 1F in particular resemble the present Solar System.  Of course, to
actually reproduce the Solar System, another important event has to
take place: The next protoplanet beyond Jupiter must also undergo a
runaway gas accretion phase to acquire an envelope of mass $\ga$
80 M$_{\oplus}$ and become Saturn.  With the innermost core on a
stable orbit in the vicinity of Saturn's present location, though---as
it is in runs 1D and 1F---the time delay between Jupiter and Saturn's
runaway phases is not strongly constrained.  A more detailed
investigation of the role of Saturn's growth will follow in Section
\ref{role_of_saturn}.  

In contrast, Uranus and Neptune must somehow have been prevented from
later reaching runaway gas accretion.  One possibility is that they
simply ran out of time, still caught in the long plateau of the second
giant planet growth phase (Pollack et al 1996; see Section
\ref{scatter+circ}) when the nebular gas was removed.  Alternatively,
it may be that the gas disk was truncated early on by
photoevaporation between the orbit of Saturn, and the eventual orbit of
Uranus (Shu, Johnstone and Hollenbach 1993).  This possibility is
discussed further in Section \ref{discussion}.

The final state of the planetesimal disk differs substantially among
the runs.  In 1B, 1C and 1E, the planetesimal disk is largely
unperturbed over most of its radial extent, while in 1F and 1H,
eccentricities have been greatly increased throughout the entire disk.
1A, 1D and 1G are intermediate cases.  The extent of the perturbation
simply depends on how much of the disk is crossed by the protoplanets
over the course of the run.  In Run 1F, for example,
Fig. \ref{run21_a6_a_peri_apo} shows that one body's aphelion spends
some time beyond 100 AU.  On the other hand, in 1C, no protoplanet's
aphelion ever goes further out than $Q_{max} \simeq$ 18 AU.  In those
runs where $Q_{max} < 60$ AU (the disk radius), $Q_{max}$ corresponds
closely to where the planetesimal disk makes a transition from
perturbed to unperturbed.  The outer limit of the planetesimal disk's
excitation by scattered protoplanets has been referred to as the
``fossilized scattered disk'' (TDL99).  The contemporary scattered
disk, by contrast, consists of objects which were scattered by Neptune
after the latter had attained its current orbit (Duncan and Levison
1997).  A more detailed discussion of the fossilized scattered disk
follows in Section \ref{kuiper belt}.

Three of the runs produce systems at 5 Myrs that are irreconcilably
different from our own.  In 1B, one of the protoplanets has merged
with Jupiter.  In 1C, a protoplanet has been scattered onto an orbit
interior to Jupiter, in the region of the present-day asteroid belt,
with its semimajor axis at 3.4 AU, its perihelion at 2.8 AU and its
aphelion at 3.9 AU.  It attains a stable orbit not crossing Jupiter
even though the only planetesimals available for damping interior to
Jupiter are those few which are also scattered there.  This is
possible because the protoplanet is scattered not just by Jupiter, but
also by the other protoplanets.  Even assuming the protoplanet could
be subsequently removed from this region, such an event would most
likely have cleared much of the asteroid belt.  Finally, in 1E, two of
the protoplanets have merged.  It should be noted that mergers---or,
indeed, ejections---which reduce the number of protoplanets are not
intrinsically a problem, since extra ones may have existed.  Scenarios
with five protoplanets (Jupiter + Saturn + 3) will be explored in
Section \ref{subset_1c}.

\subsection{Set 2: Dependence on initial ordering}

How strongly does the final configuration of the system depend on the
initial ordering of the protoplanets?  The next set of runs, Set 2,
uses the same initial conditions, within a random variation in the
protoplanet phases, as Set 1.  However, for these runs it is the
second-innermost protoplanet, rather than the innermost one, which
undergoes simulated runaway gas accretion.  One can reasonably expect that
this will favour an outcome like 1C (Fig. \ref{run21_a_ea_thesis}),
where a protoplanet is scattered inward into the region of the
asteroid belt.  
\begin{figure}
\begin{center}
\includegraphics[width=5.0in]{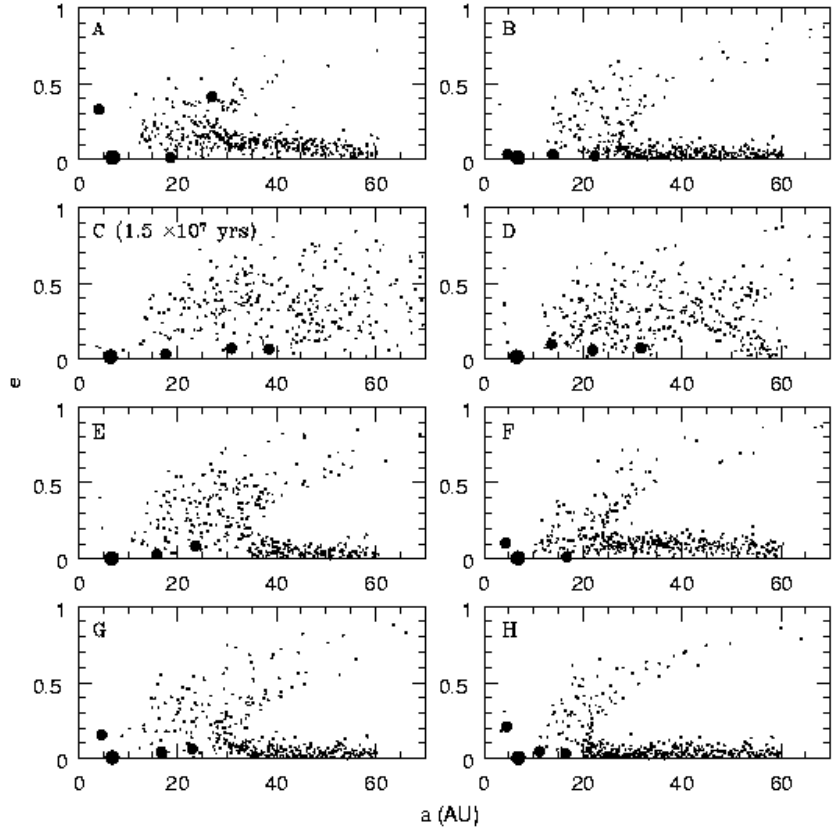}
\caption{End states of the eight Set 2 runs, after 5 Myrs of
simulation time, except for 2C, which was continued on to 15 Myrs.
Eccentricity is plotted versus semimajor axis.  The three different
sizes of points denote planetesimals (smallest), 10 M$_{\oplus}$
protoplanets (medium), and Jupiter (largest).}
\label{run21_e_ea_thesis}
\end{center}
\end{figure}

The end states of the runs are shown in Fig. \ref{run21_e_ea_thesis}.
In six of the eight runs, a protoplanet has indeed ended up interior
to Jupiter.  However, in two cases (2C and 2D), Jupiter is the
innermost body.  Thus it appears that if Jupiter does not grow from
the innermost protoplanet, the likelihood of ending up with a final
configuration similar to our Solar System declines, though such an
outcome continues to be quite possible.  

\subsection{Set 3: Dependence on number of cores}
\label{subset_1c}

\begin{figure}
\begin{center}
\includegraphics[width=5.0in]{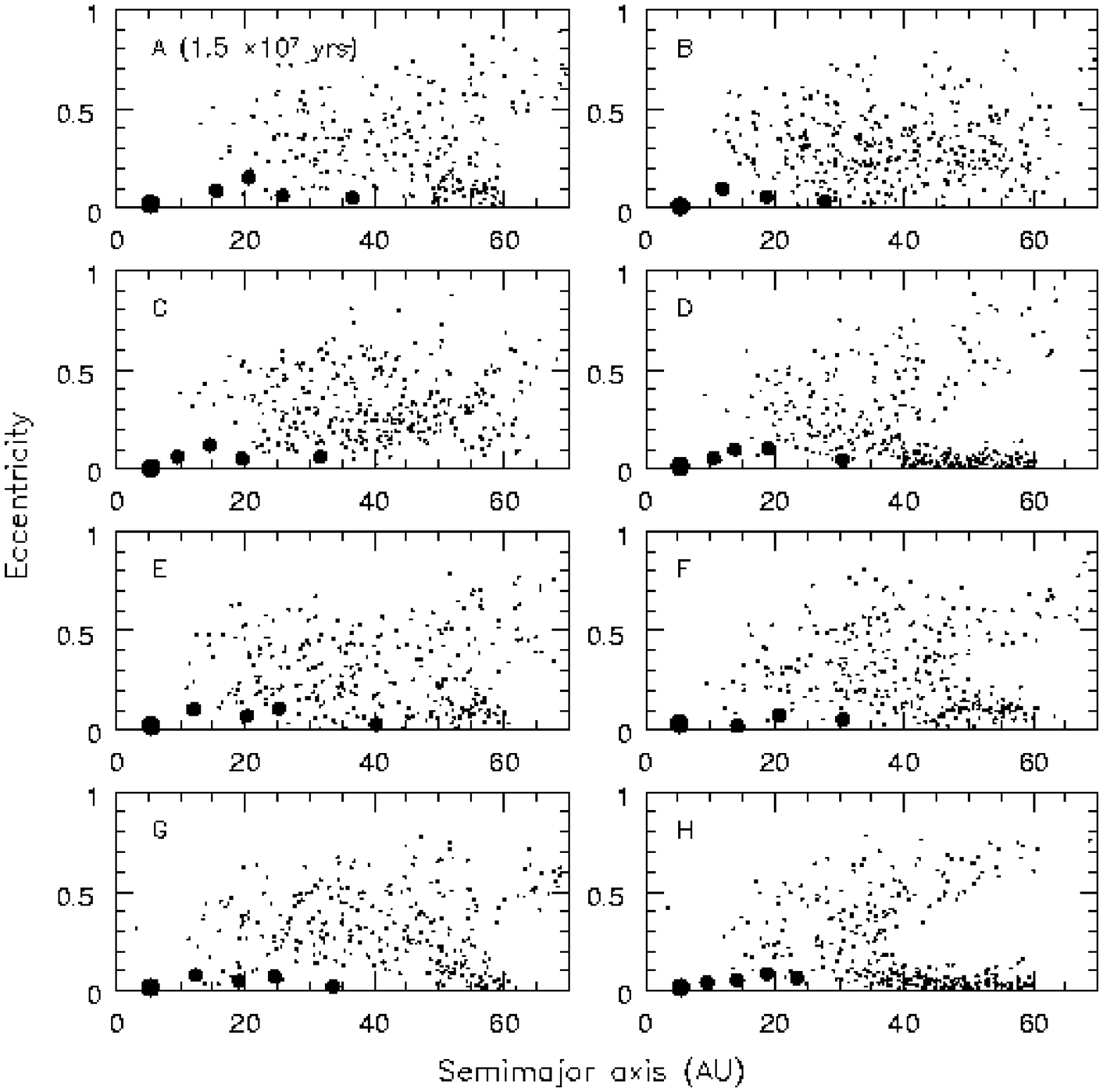}
\caption{End states of the eight Set 3 runs, after 10 Myrs of
simulation time, except for 3A, which was continued on to 15 Myrs.}
\label{ea_p}
\end{center}
\end{figure}

How sensitively does the end state depend on the initial number of
core-sized bodies?  In the next set of simulations, an extra
10 M$_{\oplus}$ protoplanet is added.  All protoplanets are more
tightly spaced, by 6.5 instead of 7.5 mutual Hill radii.  Starting,
again, from 6.0 AU, the outermost protoplanet is therefore
initially at 12.2 AU.  The surface density of planetesimals in the
region of the protoplanets is reduced to keep the average surface
density unchanged at $\sigma_1 = 10(a/5\,\mbox{AU})^{-2}$ g/cm$^2$.  The
innermost protoplanet is, again, increased in mass to that of Jupiter
over the first $10^5$ years.  

Fig. \ref{ea_p} shows the endstates of the runs.  This time all runs
initially have a length of $10^7$ years, since the larger number of
bodies take longer to decouple from each other.  Run 3A is continued
to $1.5 \times 10^7$ years, because after $10^7$ years some of the
protoplanets are still on crossing orbits.  Eccentricities are
uniformly low and the protoplanet orbits are well-spaced for the most
part, thus the systems have a good chance of being stable
indefinitely.  However, once all the planetesimals have been scattered
from among the protoplanets, so that the latter are no longer subject
to dissipative forces, some of these systems may still become
unstable.  This caveat applies to all the runs presented in this
work, but generally speaking, larger numbers of bodies increase the
potential for instability (Levison, Lissauer and Duncan 1998). 

All of the protoplanets remain in six of the eight runs, thus
resulting in systems with one too many planets relative to the present
Solar System.  However, in Runs 3B and 3F, one of the protoplanets is
ejected from the Solar System, leaving the right number of bodies
behind.  In both of these runs, a protoplanet ends up with a semimajor
axis within $10 \%$ of present-day Uranus and Neptune, respectively,
though both ``Saturns'' are too far out.  One may conclude that with
one extra initial protoplanet in the Jupiter-Saturn region, scattered
protoplanets continue to be readily circularized, and the resulting
systems tend to look like ours with one extra outer planet.  However,
a system with four giant planets remains a possible outcome.  Also,
the subsequent formation of Saturn from one of the protoplanets may
trigger more ejections, especially in those cases where the inner
protoplanets are more closely spaced, such as 3D and 3H.

\subsection{Set 4: A more massive planetesimal disk}
\label{set_2}

\begin{figure}
\begin{center}
\includegraphics[width=5.0in]{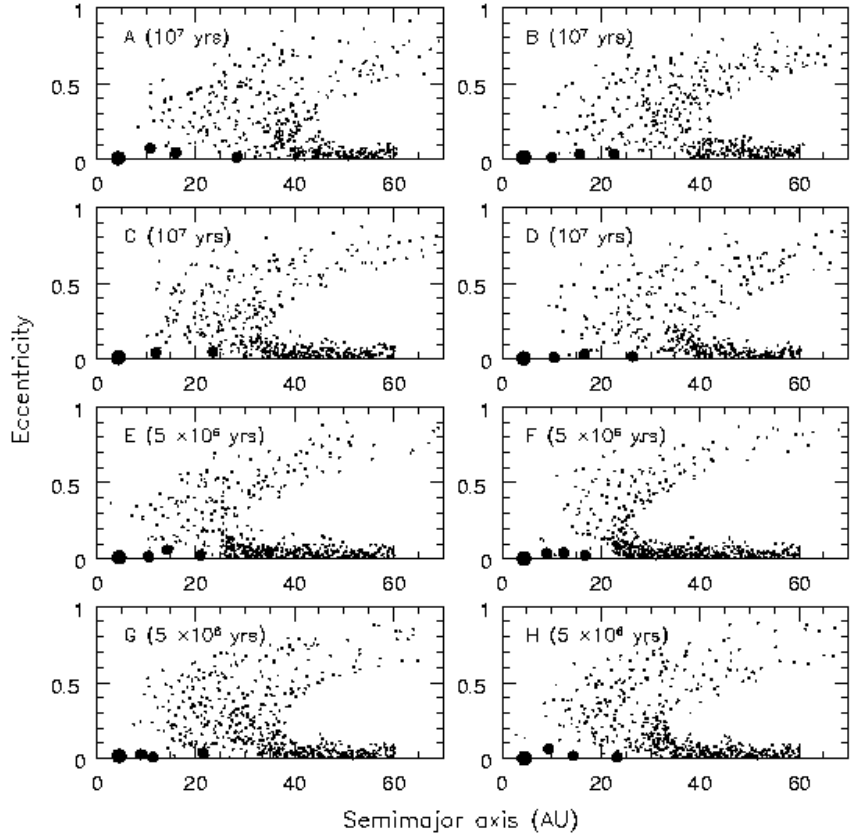}
\caption{End states of the eight Set 4 runs.  The first four are run
to $5 \times 10^6$ years; the last four are run to $10^7$ years.}
\label{run13_ea_thesis}
\end{center}
\end{figure}

In this set of runs, a number of parameters are changed to simulate a
system with a more massive planetesimals disk.  The protoplanets are
now 15 M$_{\oplus}$ bodies.  The planetesimal disk surface density
profile is still $\propto a^{-2}$, but it is scaled up to be 15
g/cm$^2$ at 5 AU; that is,
\begin{equation}
\sigma_2 = 15(a/5\mbox{ AU})^{-2}\,g/cm^2.
\label{density_set2}
\end{equation}
Other minor differences are a legacy of chronologically earlier runs.
The innermost protoplanet is initially at 5.3 AU, and successive
protoplanets are spaced by only 5.8 mutual Hill radii.  Thus the
outermost protoplanet is initially at 9.0 AU.  The individual
planetesimals have a mass of 0.24 M$_{\oplus}$.

The innermost protoplanet, as before, has its mass increased to 314
M$_{\oplus}$ over the first $10^5$ simulation years.  The end states
of the runs are shown in Fig. \ref{run13_ea_thesis}.  All except C
yield the correct number and ordering of bodies, and eccentricities
are uniformly low.  This ``success rate'' is higher than that of Set
1, in which only five out of eight runs yield qualitatively the
correct orbital configuration.  This is accounted for by the more
massive planetesimal disk; it provides stronger dynamical friction, so
that scatterings of protoplanets tend to be less violent, and
subsequently, orbits tend to be circularized and mutually decoupled
more quickly.  Runs A, D and H end up with protoplanet orbits that are
particularly close to those of Saturn, Uranus and Neptune.  Jupiter
systematically ends up at too small a heliocentric distance,
indicating that the initial distance of 5.3 AU is too small.  Also,
the larger disk mass gives the protoplanets and Jupiter more
planetesimals to scatter, and thus increases the distance they travel
due to angular momentum exchange (Section \ref{set 1}).

\subsection{Set 5: A shallower disk density profile}
\label{set_3}
\begin{figure}
\begin{center}
\includegraphics[width=5.0in]{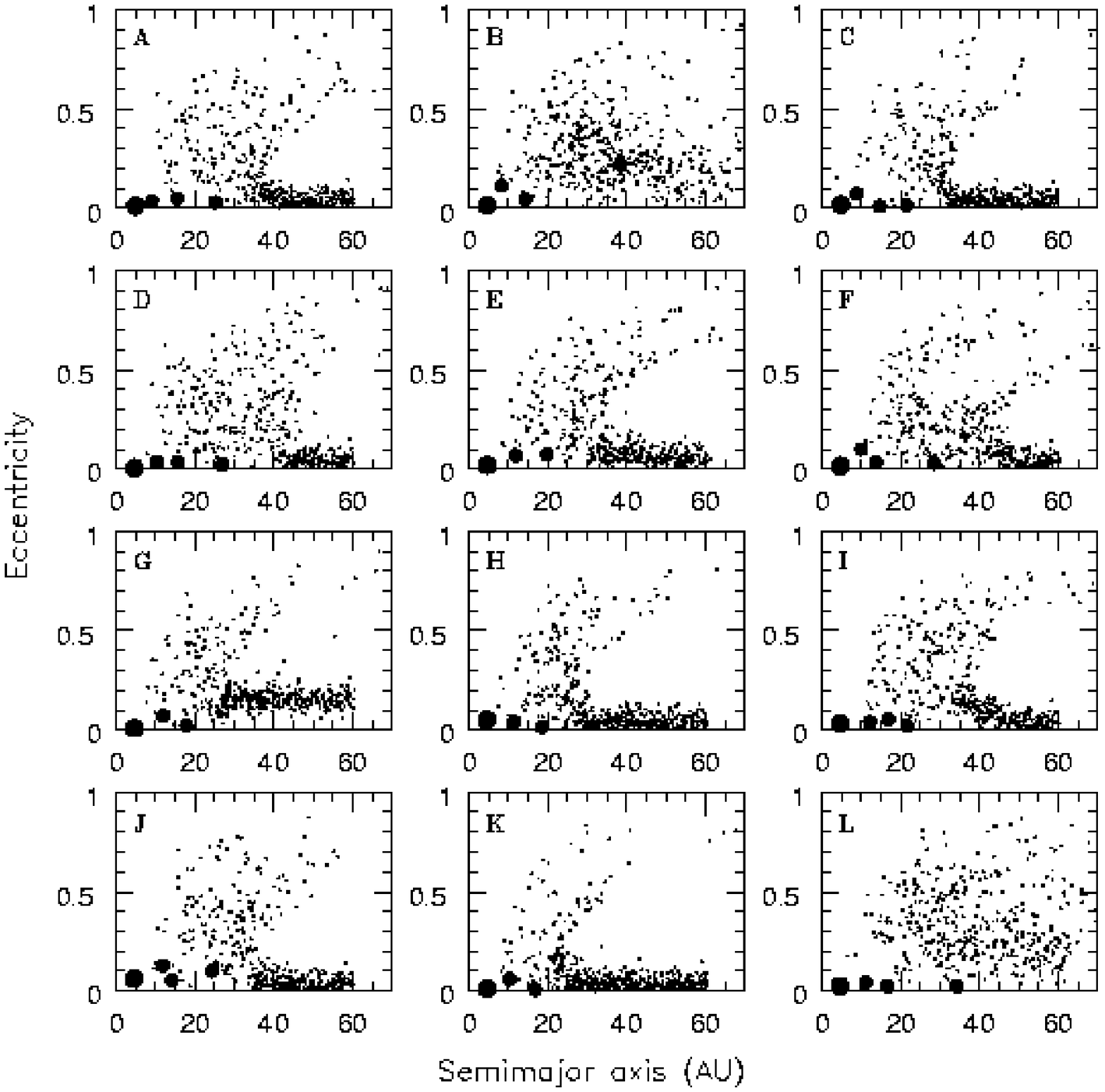}
\caption{End states of the twelve Set 5 runs.  All are run to $5
\times 10^6$ years.}
\label{run11_ea_thesis}
\end{center}
\end{figure}

In this set of runs, a shallower planetesimal disk surface density,
$\propto a^{-3/2}$, is used.  The disk now begins at 10 AU, with
the interior region being initially occupied solely by the
protoplanets.  In other words, it is assumed that in the epoch from
which the runs start, all but a negligible mass of the planetesimals
among the protoplanets has been swept up or scattered from the
region.  The surface density profile is given by
\begin{equation}
\sigma_3 = 1.8 (a/10 \mbox{ AU})^{-3/2}\,g/cm^2
\label{density_set3}
\end{equation}
and thus a total mass of 123 M$_{\oplus}$ is contained in the disk
between 10 and 60 AU.  Extending this planetesimal disk inward to 5 AU
would yield a surface density there of only 5 g/cm$^2$, and a total
mass between 5 and 10 AU of only 25 M$_{\oplus}$.  Despite this, we
still put four 15 M$_{\oplus}$ in this region; thus, it is assumed
that the original planetesimal surface density profile in this region
was steeper, perhaps due in part to redistribution of water vapor from
other parts of the disk to the vicinity of the snow line (Stevenson
\& Lunine 1988).

This set consists of twelve runs, each to 5 Myrs.  As before, the
inner body's mass is increased to 314 M$_{\oplus}$ over the first
$10^5$ years of simulation time.  The endstates are shown in
Fig. \ref{run11_ea_thesis}.  Eight out of twelve runs possess the
right number and ordering of bodies.  Eccentricities of the
protoplanets and Jupiter are $\la 0.1$ in five of these.  This
``success rate'' is close to that of Set 1 (which has a similar total
disk mass); this model thus does not appear to be highly sensitive to
changes in density profile alone.

Various degrees of disruption of the planetesimal disk can again be
seen.  Most show a sharp transition between a disrupted region crossed
by the scattered planetesimals, and a largely undisturbed outer
region.  In Run 5A, for example, this transition occurs at slightly
below 40AU, while in Run 5F, it is located between 45 and 50 AU.
Runs 5B and 5L show strong disruption throughout the entire disk,
indicating that all of it was crossed by one or more protoplanets.
Another state can be seen in Run 5G, where all eccentricities in the
outer part of the belt are uniformly raised.  This occurs when a
protoplanet crosses the outer disk with an inclination high enough
that it spends most of its orbit above or below, rather than inside,
the disk.  The disk planetesimals are then excited primarily by
long-range secular effects rather than by short-range scattering
encounters (TDL99).  Another example of
this effect can be seen in Run 1G above.

\subsection{Set 6: The role of Saturn}
\label{role_of_saturn}

Thus far, the only gas giant in the simulations has been Jupiter.  The
innermost of the scattered protoplanets does tend to end up near the
present location of Saturn.  However, to reproduce the Solar System,
it must at some point accrete $\sim$ 80 M$_{\oplus}$ of nebular gas.
We therefore investigate what effect the subsequent growth of a
Saturn-mass object has on our model.

A Set 1 run which produced a good Solar System analogue (1F;
Fig. \ref{run21_a6_a_peri_apo}) is used as a starting point.  At about
1 Myr, the protoplanets and Jupiter are no longer on crossing orbits.
At this time, the initially outermost protoplanet (plotted in blue)
has become the innermost one, closest to Jupiter at $\sim$ 10 AU.  We
perform a set of five runs which branch off from this point.  In these
runs, the innermost protoplanet has its mass increased to that of
Saturn over a $10^5$ year interval, starting at 1, 1.2, 1.4, 1.5 and
1.6 Myrs, respectively.  We choose to start only after the
protoplanets are on noncrossing orbits in order to avoid ending up
with an eccentric ``Saturn''; the eccentricity evolution of an
initially eccentric giant planet in a gas disk is uncertain (Lin et al
2000).

\begin{figure}
\begin{center}
\includegraphics[width=5.0in]{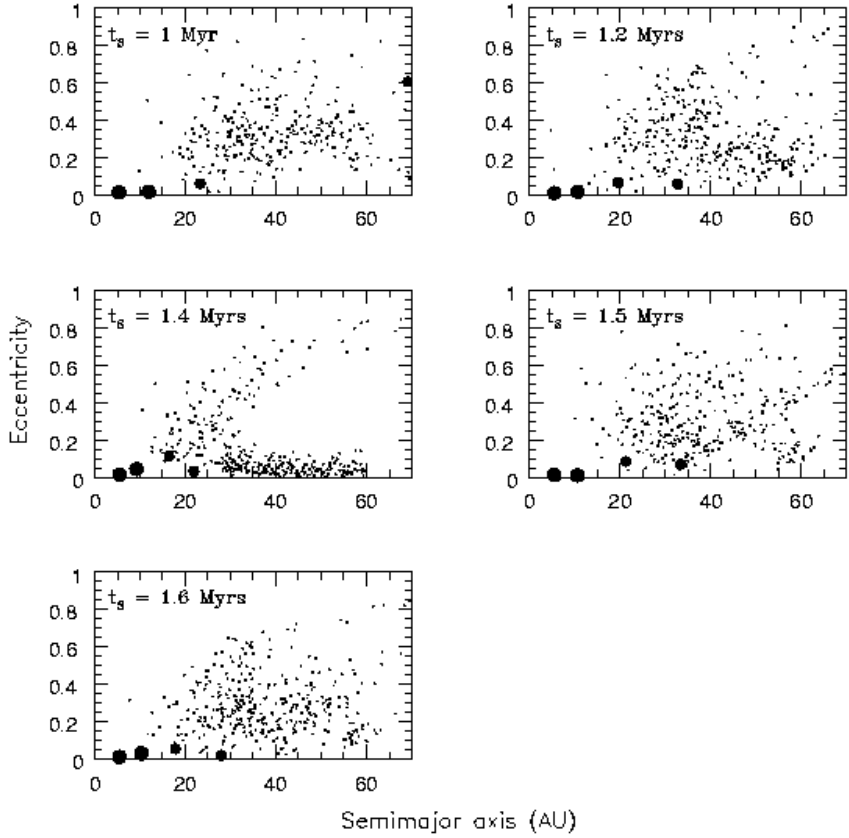}
\caption{Endstates of those Set 5 runs in which Saturn commences
growing after the protoplanets have largely decoupled from each other.
In each case, Saturn is the second-innermost of the largest points, the
innermost being Jupiter.  The start time of Saturn's growth, $t_S$, is
denoted on each panel.}
\label{ea_saturn_late}
\end{center}
\end{figure}

The endstates are shown in Fig. \ref{ea_saturn_late}.
No protoplanets have been lost from the system by the end of the runs.
A protoplanet still has a high eccentricity in the 1 Myr case;
this is because this protoplanet's perihelion is still very close to
the innermost protoplanet's orbit at 1 Myr, and it suffers strong
perturbations as the latter grows to Saturn's mass.  In the other
cases, however, Saturn's formation does not cause large eccentricities
in the protoplanets.  This is as one would expect; at its final mass,
and at 10 AU, Saturn's Hill radius is 0.45 AU, and by $1.2 \times
10^6$ years, the closest protoplanet's perihelion is at $\sim$ 14 AU,
almost $9\,r_H$ away, thus unlikely to be in reach of strong
scattering.  We leave for future work the effect of Saturn's gas
accretion on systems with more protoplanets, such as those in Set 3 in
Section \ref{subset_1c} above.  Such systems tend to have weaker
stability, and thus may be more susceptible to disruption by Saturn's
final growth spurt.   

One effect visible in Fig. \ref{ea_saturn_late} is that the semimajor
axis of Saturn at 5 Myrs tends to be smaller than that of the
innermost scattered protoplanet---the putative proto-Saturn---in those
runs where the gas accretion of Saturn is not modeled (see for example
Fig. \ref{run21_a_ea_thesis}).  When a protoplanet grows to Saturn's
mass, its subsequent migration speed is much slower, since the rate of
migration depends on how much mass in planetesimals it scatters
relative to its own mass.  This counteracts the tendency of the
innermost scattered protoplanet to end up at a semimajor axis larger
than that of present-day Saturn in the other runs, where only Jupiter
grows.

\section{Scattered protoplanets and the small body belts}
\label{small body belts}
The simulations presented above show that scattering of giant planet
core-sized protoplanets is a violent event, which leaves a strong
dynamical signature on the surrounding planetesimal disk.  The
asteroid belt and the Kuiper belt are therefore the natural places to
look for evidence of large scattering events in the Solar
System's early history.

\subsection{The asteroid belt}
\label{asteroid belt}

It is those members of the asteroid belt larger than about 50 km in
diameter which are of interest in inferring properties of the early
Solar System; smaller bodies cannot be primordial because they could
not have survived intact for the age of the Solar System (eg. Petit,
Morbidelli and Valsecchi 1999).  This population displays a degree of
dynamical excitation that is not readily explained.  The other
puzzling feature of the asteroid belt is its severe mass
depletion---by at the very least a factor of $10^3$---relative to the
amount of mass the region originally contained, extrapolated from the
terrestrial zone (Weidenschilling 1977).

\begin{figure}
\begin{center}
\includegraphics[width=5.0in]{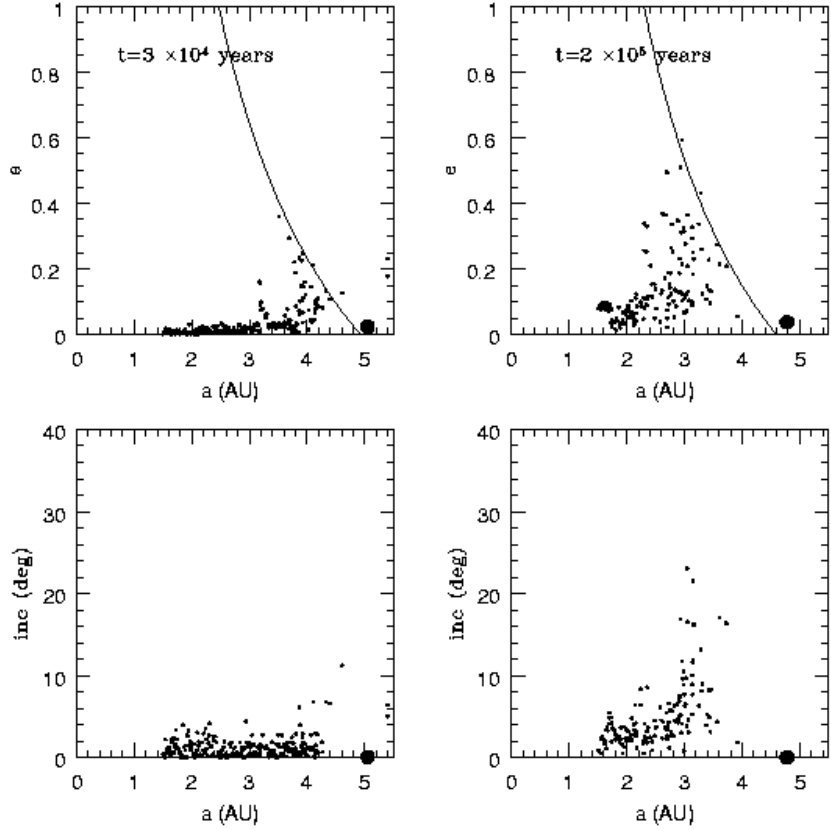}
\caption{Eccentricities and inclinations of planetesimals in the
asteroid belt region, interior to Jupiter (the large dot), at $3 \times
10^4$ years (top panel) and $2 \times 10^5$ years (bottom panel).
These times are, respectively, just before and just after the period
during which a protoplanet repeatedly crossed the region interior to
Jupiter.  Jupiter in this run
has moved inward to $\sim$ 4.8 AU, 0.4 AU less than its present
semimajor axis.  The curve marks the locus of Jupiter-crossing
orbits}
\label{run16_eia}
\end{center}
\end{figure}

In the context of the model presented here
for the early evolution of Uranus and Neptune, the initial violent
scattering of protoplanets is perhaps the most obvious candidate to
look to for perturbation of the asteroid belt.  Indeed, in numerous
runs, protoplanets spend some time interior to the orbit of Jupiter,
crossing part of the asteroid belt region for up to $\sim 10^4$ years
at a time.  One might expect that such an occurence would wreak havoc
in the the asteroid belt.  To explore this possibility, we perform six
additional runs with planetesimals added in the asteroid belt region.
The individual bodies have a mass of $0.024\,M_{\oplus}$, and are
distributed with a surface density of
\begin{equation}
\sigma_{belt} = 8.0 (a/1 \mbox{ AU})^{-1}\,g/cm^2
\label{belt_sigma}
\end{equation}
between 2.5 and $4.5\,AU$.  This shallow density profile is the same
as the one used by Chambers and Wetherill (1998) in the terrestrial
region, which in turn was chosen to be more consistent with the large
densities required at larger heliocentric distances to form Jupiter
and Saturn.  The protoplanet masses in this case are $15\,M_{\oplus}$.
We find that eccentricities can get excited to their present values in
this way, though only down to the crossing protoplanet's minimum
perihelion distance, which in none of the runs performed reaches the
inner edge of the belt.  Inclinations fare more poorly; protoplanets
seldom raise them much above $10 \degr$, which is less than the
present median inclination of asteroids beyond 2.5 AU.  Also, very
little mass is scattered out of the belt while the protoplanets are
crossing it.  Fig. \ref{run16_eia} shows ``before and after''
snapshots of the run which, out of the six, displays the strongest
disruption of the region interior to Jupiter by scattered
protoplanets.  As can be seen, few bodies, apart from those that are
nearly Jupiter-crossing, attain inclinations above $15\degr$.

Scattered protoplanets may have had a more indirect role in the
excitation and mass depletion of the asteroid belt.  If the majority
of the solids in the asteroid belt accreted to form a system of
planet-sized bodies (raising eccentricities and inclinations of the
remaining planetesimals in the process), scattered protoplanets
crossing into the region may have contributed to making this system
unstable, akin to the model of Chambers (1999).  Also, if proto-Saturn
was still sufficiently close to Jupiter when it accreted its massive
gas envelope, the subsequent migration of the gas giants may have been
enough to sweep the inclination-exciting $\nu_{16}$ secular resonance
through most of the asteroid belt (cf. Gomes 1997, Levison et al 2001)
As an example, in one of the runs from Set 6, Saturn commences growing
at $4 \times 10^5$ years and reaches its final mass at $5 \times 10^5$
years, at which point Jupiter and Saturn are at 5.7 and 8.4 AU,
respectively.  This places the $\nu_{16}$ resonance at $\sim 3.5$ AU,
and it moves inward as Jupiter and Saturn move apart.  In contrast,
Gomes (1997), using a more moderate range of migration for Jupiter and
Saturn, finds that only the region inward of 2.7 AU is crossed by the
$\nu_{16}$ resonance.

\subsection{The Kuiper belt}
\label{kuiper belt}

In the present Solar System, a new class of Kuiper belt object (KBO)
has recently been identified (Duncan and Levison 1997, Luu et al
1997).  These objects have semimajor axes and eccentricities such that
they lie near the locus of Neptune-encountering objects shown in
Fig. \ref{real_solar_system}.  They are thought to be part of a
population referred to collectively as the scattered disk---formerly
low-eccentricity KBOs which have had their orbits changed by close
encounters with Neptune.  Many of the simulations in Section
\ref{n-body simulations} show an analogous class of planetesimals in
their ``Kuiper belt'' regions.  However, these fall on the locus of
orbits crossing not the final semimajor axis of the outermost
protoplanet, but the furthest aphelion distance of any of the
protoplanets during their initial high-eccentricity phase.  Since
these orbits are no longer being crossed by a protoplanet, they will
be stable over long times.  One can refer to these structures as
``fossilized'' scattered disks (TDL99), because they preserve part of
the dynamical history of the planetesimal disk.  Such structures only
appear in runs where the initial scattering was strong enough that one
or more protoplanets had their aphelia increased to well beyond the
final semimajor axis of the (ultimately) outermost protoplanet.

Observations of our Solar System's Kuiper belt do indeed reveal an
anomalously high degree of excitation (eg. Petit, Morbidelli \&
Valsecchi 1999, Malhotra, Duncan \& Levison 2000).  The
eccentricities and, to a lesser degree, inclinations of bodies in
mean-motion resonances with Neptune, particularly the 2:3 resonance at
39.5 AU, can be explained by resonance sweeping during Neptune's
migration, as can the paucity of objects on nonresonant orbits
interior to 39 AU (Malhotra 1995).  However, the high inclinations
found beyond $\sim$ 41 AU in what is commonly called the ``classical''
Kuiper belt, cannot be explained in this way.  Petit, Morbidelli and
Valsecchi (1999) propose large (up to 1 M$_{\oplus}$)
Neptune-scattered planetesimals as the mechanism which stirred and
cleared the belt.  But even when such bodies remain in the belt for
100 Myrs, the inclinations they raise are almost always less than
$20\degr$.

\begin{figure}
\begin{center}
\includegraphics[width=5.0in]{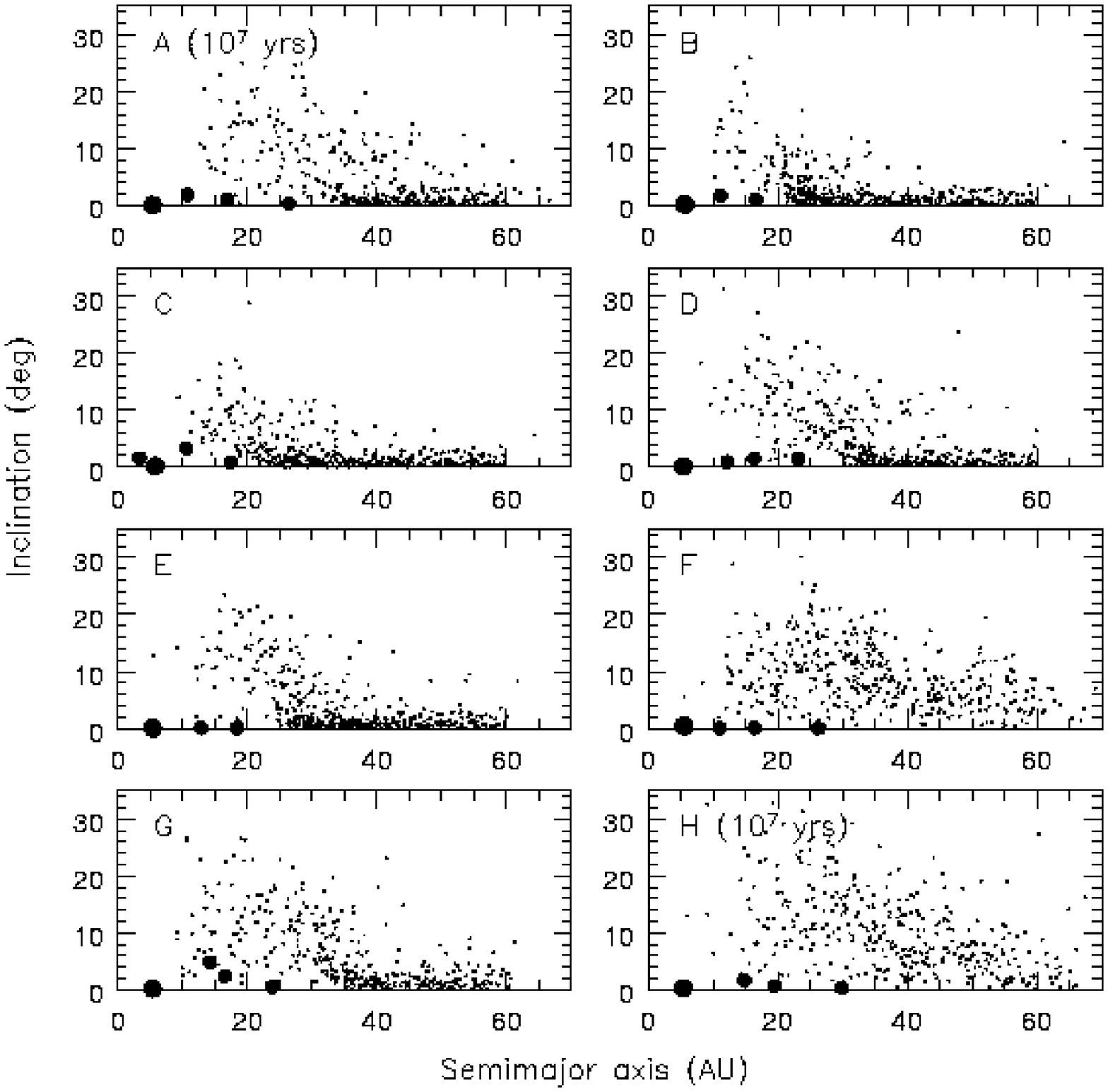}
\caption{Counterpart to Fig. \ref{run21_a_ea_thesis}, showing
inclination versus semimajor axis for the endstates of the runs in
Set 1.}
\label{ia_a_thesis}
\end{center}
\end{figure}

\begin{figure}
\begin{center}
\includegraphics[width=5.0in]{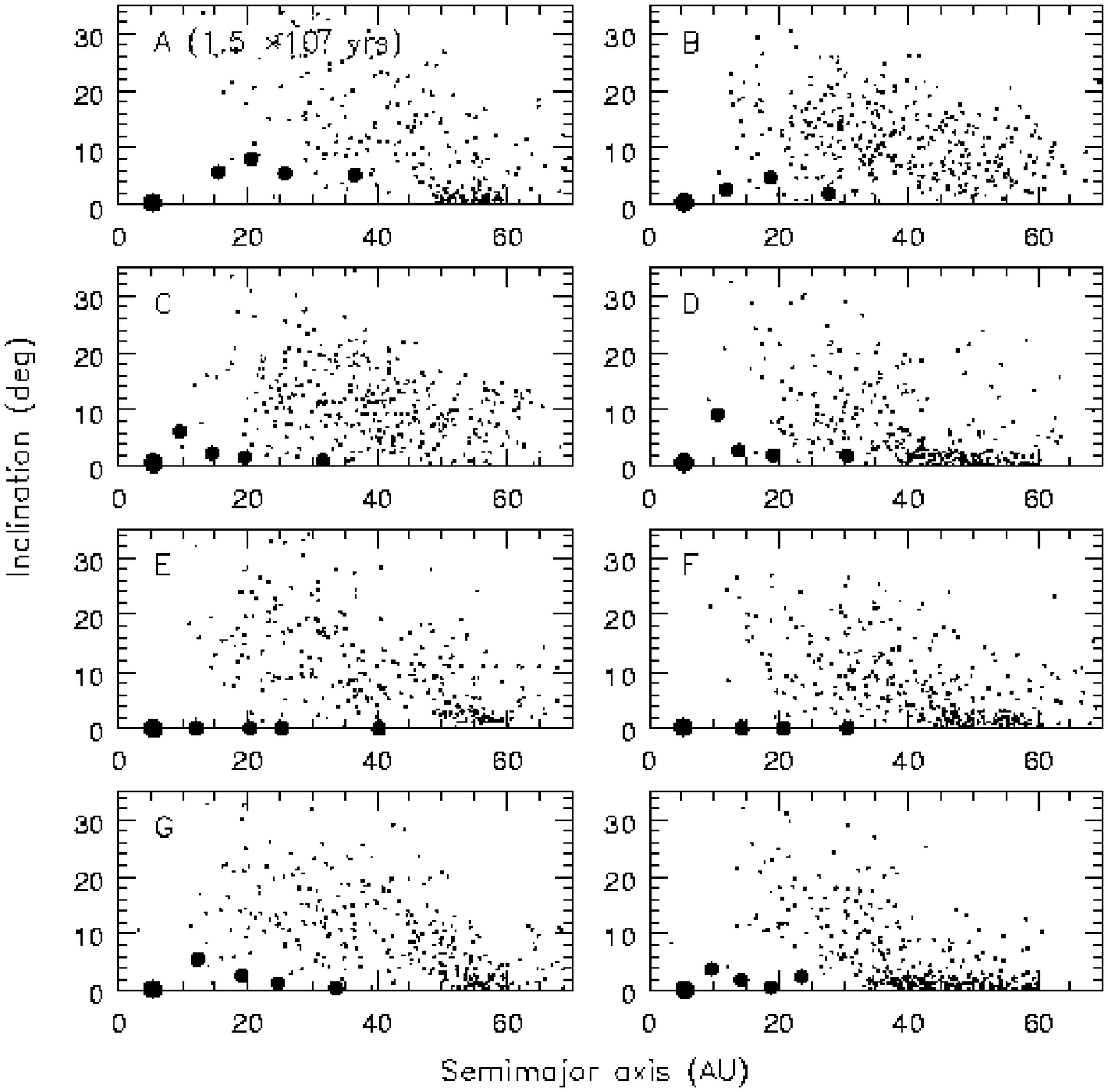}
\caption{Counterpart to Fig. \ref{ea_p}, showing
inclination versus semimajor axis for the endstates of the runs in
Set 3.}
\label{ia_p_thesis}
\end{center}
\end{figure}

\begin{figure}
\begin{center}
\includegraphics[width=5.0in]{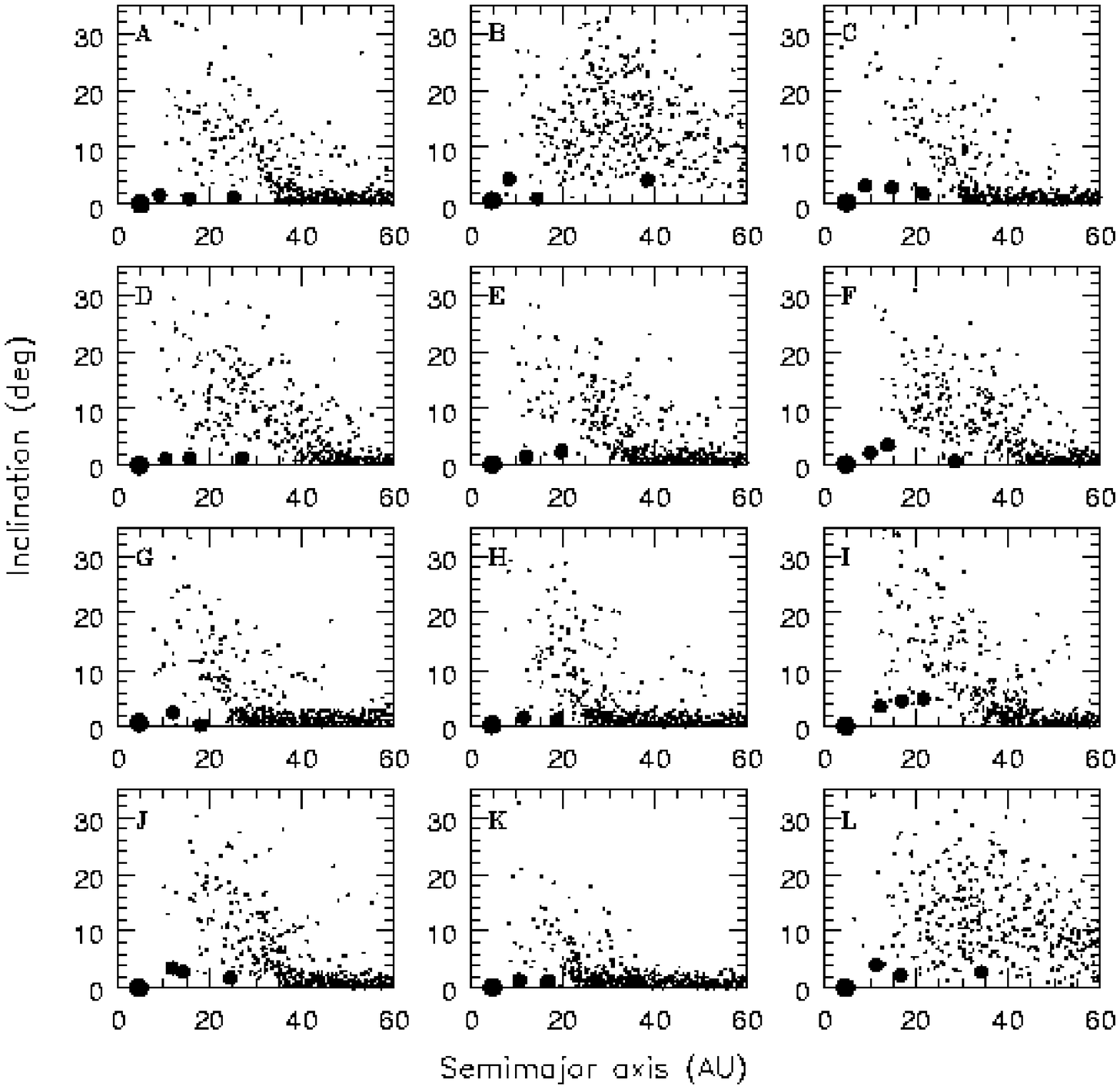}
\caption{Counterpart to Fig. \ref{run11_ea_thesis}, showing
inclination versus semimajor axis for the endstates of the runs in Set
5.}
\label{ia_run11}
\end{center}
\end{figure}

Can the excitation of the Kuiper belt be better accounted for if we
are thus far only seeing the part of it that is interior to the locus
of a fossilized scattered disk?  The inclinations raised by a
protoplanet can be directly obtained from the simulations of Section
\ref{n-body simulations}.  The runs of Set 1, Set 3 and Set 5
will be used for comparison.  Inclinations for Sets 1, 3 and 5 are
shown in Figs. \ref{ia_a_thesis}, \ref{ia_p_thesis} and
\ref{ia_run11}, respectively.  Inclinations beyond the outermost
protoplanet are excited up to a maximum of about $30 \degr$, similar to
those observed in the classical Kuiper belt today (see
Fig. \ref{real_solar_system}) However, observations of the Kuiper belt
are biased against high-inclination objects.  Brown (2001) derives a
de-biased inclination distribution function for the classical belt:
\begin{equation}
f_t(i)=\sin(i) \left [ a\, \exp \left ( \frac{-i^2}{2 \sigma_1^2} \right )
+ (1-a) \exp \left  ( \frac{-i^2}{2 \sigma_2^2} \right ) \right ]
\label{debiased inclination}
\end{equation}  
with $a=0.93 \pm 0.02$, $\sigma_1=2.2 \pm^{.2}_{.6}$, and $\sigma_2=17 \pm 3$.  
One can define a parameter $i' \equiv \cos^{-1}(\overline{\cos(i)})$
to give a measure of the characteristic inclination of a population.
For the de-biased distribution above, $i'=21\degr$.  However, in the
runs presented here, the largest $i'$ in the region corresponding to
the classical belt (between the outermost planet's 2:3 and 1:2
mean-motion resonance) is only $15\degr$.  Thus, although higher
inclinations are produced here than in the large Neptune-scattered
planetesimals model of Petit, Morbidelli and Valsecchi (1999), the
inferred full velocity distribution of the classical Kuiper belt still
cannot be accounted for.  

\begin{figure}
\begin{center}
\includegraphics[width=5.0in]{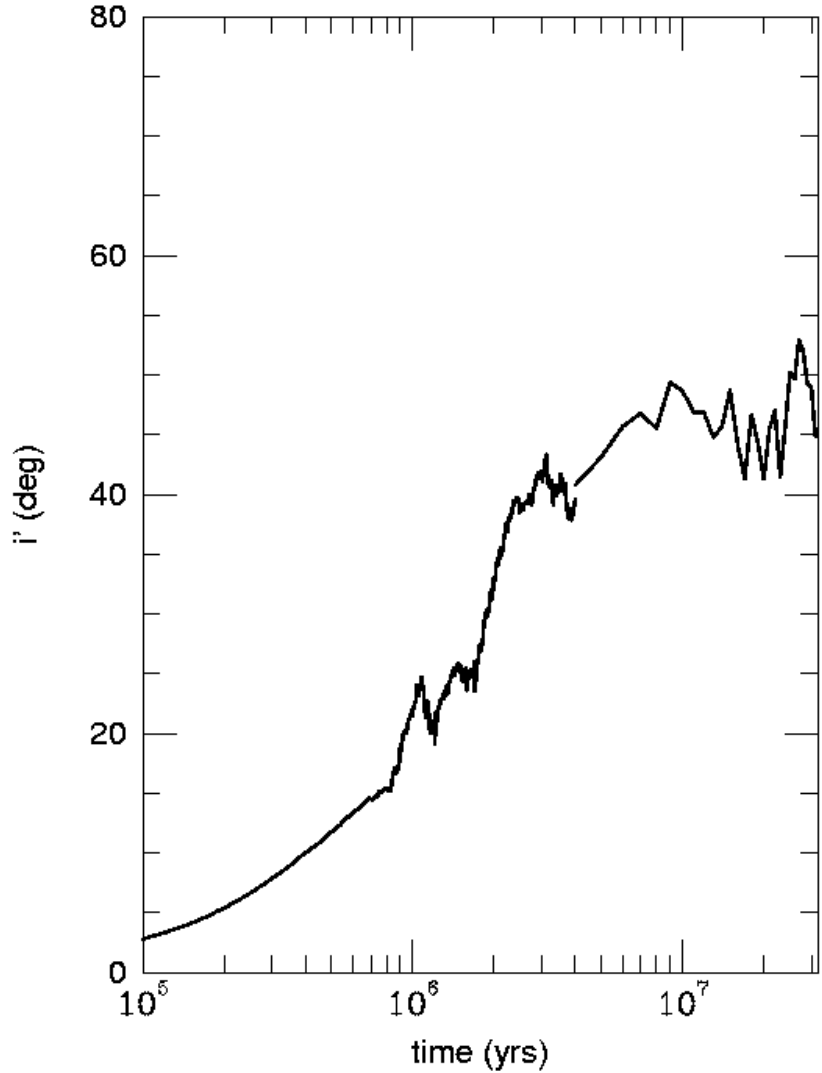}
\caption{Characteristic inclination $i'$ versus time of test
particles between 35 and 55 AU, gravitationally stirred by a
Uranus-mass body initially having $a=45$ AU, $e=0.25$, and
$i=25\degr$.  The test particle inclinations reach the debiased value
for the Kuiper belt, $i' = 20\degr$ (Brown 2001), after about 1 Myr.}
\label{iprime}
\end{center}
\end{figure}

It is possible to estimate with a simple numerical experiment if a
planet as large as Uranus or Neptune can in principal excite the
Kuiper belt to observed values.  This experiment consists of a single
Uranus-mass planet on an orbit with $a=45AU$, $e=0.2$, and
$i=25\degr$.  The planet is embedded in a swarm of 400 massless test
particles informally spread from 35AU to 55AU, with initial $e=0.01$
and $i=1\degr$.  Fig. \ref{iprime} plots the $i'$ of the particles as
a result of scattering off of the planet.  It shows that a planet the
mass of the ice giants can indeed excite the Kuiper belt to
$i'=20^\circ$ in a million years.

However, in none of the runs we performed was such a high inclination
imparted on a protoplanet.  Alternatively, a less inclined protoplanet
may be able to reproduce the Kuiper belt inclinations if it remains on
a belt-crossing orbit significantly longer than 1 Myr (and thus longer
than in our runs).  This may require a less massive trans-Saturnian
planetesimal disk, in order to increase the dynamical friction
timescale, though that in turn would increase the chances of
protoplanets being scattered out of the Solar System altogether.  A
longer circularization time will also result if a protoplanet can be
decoupled from Jupiter and the other protoplanets at a larger
semimajor axis, so that the planetesimal surface density averaged over
its orbit is lower.  For the latter situation to have a better chance
of occurring, the planetesimal disk needs to be extended to larger
heliocentric distance.  We will investigate this issue further in
future work.

\section{Discussion}
\label{discussion}

The conventional picture of Uranus and Neptune's formation, whereby
the ice giants accrete near their present heliocentric distances, has
grave problems.  Numerical simulations have not been able to produce
$\sim$ 10 M$_{\oplus}$ objects in the trans-Saturnian region in the
lifetime of the Solar System without significantly increasing
protoplanet radii, or invoking dissipational forces and planetesimal
disk densities too large to be consistent with a physically plausible
protostellar disk.

Building on our previous work (TDL99), we have performed additional
N-body simulations of the evolution of the outer Solar System starting
at the time when the first gas giant forms.  At this point, we assume
that a number of $\sim$ 10 M$_{\oplus}$ objects have formed at a
heliocentric distance of roughly 5 to 10 AU, as is suggested by the
oligarchic growth model (Kokubo \& Ida 1998, 2000).  Using a variety
of different initial conditions, we find as before that the accretion
of Jupiter's gas envelope causes the remaining protoplanets to become
violently unstable.  In most cases they are scattered onto
high-eccentricity orbits in the trans-Saturnian region.  With most of
its orbit now crossing the largely pristine trans-Saturnian
planetesimal disk, a scattered protoplanet experiences dynamical
friction and has its eccentricity rapidly damped.  As a result, the
protoplanets tend to end up on nearly circular, well-spaced orbits on
a Myr timescale, with semimajor axes comparable to those of Saturn,
Uranus and Neptune.  Of the simulations which initially contain a
total of four giant protoplanets and form Jupiter from the innermost,
the majority produce final orbital configurations similar to that of
our outer Solar System.  Such systems are produced---though with lower
probability---even if ones adds an extra protoplanet, or lets the
second-innermost protoplanet grow into Jupiter.  These results
strengthen our earlier conclusion that if Uranus and Neptune shared
the same birthplace as the gas giants, they could then readily have
been delivered to their present orbits.

The role of migration in the formation of Uranus and Neptune was
previously investigated numerically by Ipatov (1991), based on an idea
by Zharkov and Kozenko (1990).  Ipatov concludes that planetary
embryos of a few M$_{\oplus}$ may have originated just outside the
orbit of Saturn, to migrate outward and later grow into Uranus and
Neptune, provided that they did not acquire high eccentricities during
this process.  In contrast, we find that Uranus and Neptune could have
originated from anywhere in the Jupiter-Saturn region, and that
initially high eccentricities---which nearby bodies naturally tend to
acquire during Jupiter's final growth phase---are in fact a powerful
mechanism for rapidly transporting them outward.  Also, the long
growth timescales in the outer Solar System suggest that Uranus and
Neptune likely already completed most of their growth in the
Jupiter-Saturn region; even with a ``head start'' of a few Earth
masses, the formation of Uranus- and Neptune-mass objects much beyond
10 AU within the age of the Solar System is unlikely (Levison and
Stewart 2001, Thommes, Duncan and Levison 2001)

\begin{figure}
\begin{center}
\includegraphics[width=5.0in]{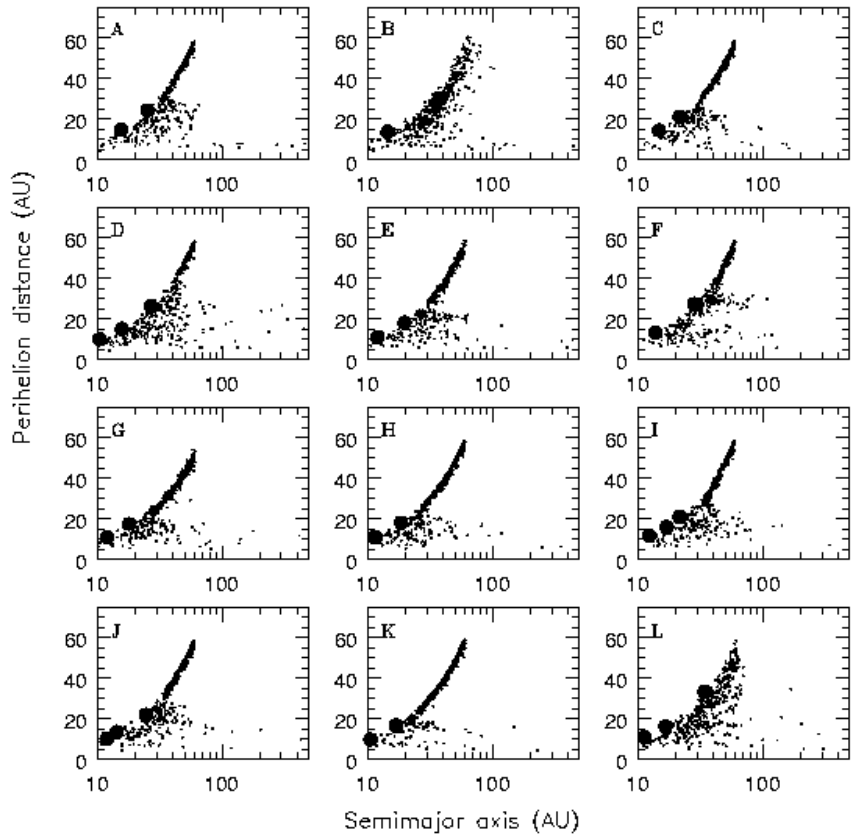}
\caption{Endstates of the Set 5 runs, showing perihelion distance
versus semimajor axis.}
\label{qa_run11}
\end{center}
\end{figure}

Can one find any observational support for this model in the
present-day Solar System?  The high inclinations in the classical
Kuiper belt point to strong dynamical excitation in the past, and the
simulations performed here do produce high inclinations in this region
as a natural side effect.  However, the simulations all fall short of
reproducing the mean debiased inclination of the classical Kuiper
belt.  Strong observational support would be provided by the discovery
of a fossilized scattered disk in the Kuiper belt, and a dynamically
colder population beyond.  It is tempting to link the trans-Neptunian
object 2000 CR$_{105}$ with a fossilized scattered disk; its high
eccentricity (0.8) is characteristic of a scattered disk object, but
recent observations (Gladman et al 2001) have established that its
perihelion is at 44 AU, far beyond the reach of Neptune.  However,
none of the fossilized disk objects in our simulations acquire
semimajor axes as high as that of 2000 CR$_{105}$ (216 AU).
Fig. \ref{qa_run11} shows a plot of perihelion distance versus
semimajor axis for Set 5, revealing only one case (B) where one or
more objects simultaneously acquire a semimajor axis of $\sim$ 100 AU
and a perihelion distance significantly further out than the
(circularized) outermost protoplanet.  All other sets of runs fare
even more poorly.  5B is a run in which a protoplanet spends a long
time at high eccentricity, and excites particularly high planetesimal
inclinations in the disk.  A long circularization time may thus be an
important ingredient in reproducing both objects like 2000 CR$_{105}$
and the high inclinations of the Kuiper belt; we will address this
possibility in future work.

Findings regarding the deuterium to hydrogen (D/H) ratios of Uranus,
Neptune and comets are particularly interesting in the context of this
model.  From infrared observations, Feuchtgruber et al (1999) find
that the D/H ratios of the ice giants are lower than those of the Oort
cloud comets Halley (Eberhardt et al 1995), Hyakutake (Bockel{\'
e}e-Morvan et al. 1998) and Hale-Bopp (Meier et al 1998) by a factor
of approximately three, a difference large compared to the
uncertainties of the measurements.  Oort cloud comets are thought to
have originated primarily in what is today the Uranus-Neptune region
(Duncan, Quinn \& Tremaine 1987, Fernandez 1997), and the notion of a
common birthplace is supported by the comets' similar D/H ratios.
Though a sample size of three is clearly very small, this discrepancy
between the comets and the ice giants would seem to present a further
problem for any scenario in which Uranus and Neptune form in the
trans-Saturnian region, since they should then share the chemical
composition of the comets.  However, this is exactly what one would
expect if the ice giants originally formed at a smaller heliocentric
distance, where higher temperatures would have made for a lower D/H
ratio.

An aspect not modeled in any of the simulations is the gravitational
interaction of the bodies with a gaseous disk.  It has been shown that
gas disk tidal forces can cause rapid inward migration of protoplanets
(eg. Ward 1997).  In fact, the speed of migration may be peaked for
objects of $\sim$ 10 M$_{\oplus}$, taking place on timescales of $10^5$ years
or less.  This peak corresponds to the transition between so-called
Type I migration, where a body's resonant interaction with the gas
disk gives rise to a torque imbalance, to Type II migration, where the
object opens a gap in the gas disk and is subequently locked to the
disk's viscous evolution.  Tidal migration therefore poses a problem
for {\it any} model of giant planet formation: how do they avoid
spiralling into the central star as they form?  However, tidal torques
may not in fact operate throughout the whole disk.  Gammie (1996)
develops an accretion disk model which, beyond $\sim$ 0.1 AU, only
transports angular momentum in a relatively thin surface layer.  Thus
the bulk of the disk would have zero viscosity, and objects embedded
in it would be subject to neither Type I nor Type II orbital decay.
Also, the disk may be truncated early on by photoevaporation from the
central star (Shu, Johnstone and Hollenbach 1993), or from surrounding
stars, as is seen in the Orion Nebula proplyds (Johnstone, Hollenbach
and Bally 1998).  Bodies scattered beyond the truncation distance
would then be safe from nebula tides, and those near the edge would
experience a net positive torque and migrate outward.  Of course, such
a scenario has the added advantage that, as discussed in the above
works, it accounts for Uranus and Neptune having no massive gas
envelopes.

In constructing the simulations presented here, we have appealed to
the oligarchic growth model as a plausible guide for our choice of
initial protoplanet masses and orbits.  One can of course envision a
number of variations.  For instance, two gas giants may form on
initially widely separated orbits.  The waves they raise in the disk
will tend to clear the gas between them, and Type II migration will
cause their orbital separation to decrease (Kley 2000).  If ice giant
sized bodies are trapped in between, they will be prevented from
accreting more gas, and will be scattered as the stable region between
the gas giants shrinks to nothing.  Also, if future measurements of
Jupiter constrain its core to be much smaller, or even absent, this
will invalidate the assumption that Jupiter grew from an ice giant
sized nucleus.  Though the scattering of ice giants could still take
place in principal, one would then need to explain how a small body
won the gas accretion race against much larger bodies (if the core is
small), or how ice giant sized bodies managed to form before the birth
of the first gas giant from an unnucleated disk instability (if there
is no core; eg. Boss 2000).  Alternatively, one could search for a way
in which accretion could continue to take place in close proximity to a
gas giant, with scattering delayed until ice giant sized bodies form.
For such a scenario Type I tidal effects, which cause the eccentricity
to decay on an even shorter timescale than the semimajor axis
(eg. Papaloizou \& Larwood 2000), may actually be helpful.  It may
then be the dispersal of the gas which triggers scattering.  Doing
away with any reference to a specific formation process, the most
general statement of our results is: {\it Ice giant sized bodies can
be scattered from the Jupiter-Saturn region by gas giant sized bodies,
to ultimately end up on Uranus- and Neptune-like orbits.}

\acknowledgements 

EWT is grateful for support from the Center for Integrative Planetary
Science, as well as from the National Sciences and Engineering
Research Council (NSERC) during the earlier part of this work.  MJD is
grateful for support from NSERC.  HFL is grateful for support from
NASA's {\it Planetary Geology \& Geophysics}, {\it Origins of Solar
Systems}, and {\it Exobiology} programs.

\end{document}